\documentclass[]{aastex631}

\begin{document}

\title{Tracking the jet-like corona of black hole Swift J1727.8-1613 during a flare state through Type-C quasi-periodic oscillations}

\correspondingauthor{Lang Cui}
\email{cuilang@xao.ac.cn}

\author[0009-0007-7562-7720]{Jie Liao}
\affiliation{Xinjiang Astronomical Observatory, CAS, 150 Science-1 Street, Urumqi 830011, China}
\affiliation{College of Astronomy and Space Science, University of Chinese Academy of Sciences, No.1 Yanqihu East Road, Beijing 101408, China}

\author[0000-0002-8684-7303]{Ning Chang}
\affiliation{Xinjiang Astronomical Observatory, CAS, 150 Science-1 Street, Urumqi 830011, China}

% \collaboration{20}{(AAS Journals Data Editors)}

\author[0000-0003-0721-5509]{Lang Cui}
\affiliation{Xinjiang Astronomical Observatory, CAS, 150 Science-1 Street, Urumqi 830011, China}
\affiliation{Key Laboratory of Radio Astronomy and Technology, CAS, A20 Datun Road, Chaoyang District, Beijing, 100101, China}
\affiliation{Xinjiang Key Laboratory of Radio Astrophysics, 150 Science 1-Street, Urumqi 830011, China}

\author[0000-0003-3166-5657]{Pengfei Jiang}
\affiliation{Xinjiang Astronomical Observatory, CAS, 150 Science-1 Street, Urumqi 830011, China}
\affiliation{Key Laboratory of Modern Astronomy and Astrophysics (Nanjing University), Ministry of Education, Nanjing 210023, China}

\author{Didong Mou}
\affiliation{School of Physics and Astronomy, Xihua Normal University, Nanchong 637009, China}
\affiliation{Xinjiang Astronomical Observatory, CAS, 150 Science-1 Street, Urumqi 830011, China}

\author[0000-0001-7199-2906]{Yongfeng Huang}
\affiliation{School of Astronomy and Space Science, Nanjing University, Nanjing 210023, China}
\affiliation{Key Laboratory of Modern Astronomy and Astrophysics (Nanjing University), Ministry of Education, Nanjing 210023, China}

\author[0000-0003-4341-0029]{Tao An}
\affiliation{Shanghai Astronomical Observatory, CAS, 80 Nandan Road, Shanghai 200030, China}

\author[0000-0001-6947-5846]{Luis C. Ho}
\affiliation{Kavli Institute for Astronomy and Astrophysics, Peking University, Beijing 100871, China}
\affiliation{Department of Astronomy, School of Physics, Peking University, Beijing 100871, China}

\author[0000-0001-7584-6236]{Hua Feng}
\affiliation{Key Laboratory of Particle Astrophysics, Institute of High Energy Physics, CAS, Beijing 100049, China}

\author{Yu-Cong Fu}
\affiliation{School of Physics and Astronomy, Beijing Normal University, Beijing, 100875, China}

\author{Hongmin Cao}
\affiliation{School of Electronic and Electrical Engineering, Shangqiu Normal University, 298 Wenhua Road, Shangqiu 476000, China}

\author[0000-0002-3960-5870]{Ashutosh Tripathi}
\affiliation{Xinjiang Astronomical Observatory, CAS, 150 Science-1 Street, Urumqi 830011, China}

\author[0000-0001-9815-2579]{Xiang Liu}
\affiliation{Xinjiang Astronomical Observatory, CAS, 150 Science-1 Street, Urumqi 830011, China}
\affiliation{Key Laboratory of Radio Astronomy and Technology, CAS, A20 Datun Road, Chaoyang District, Beijing, 100101, China}
\affiliation{Xinjiang Key Laboratory of Radio Astrophysics, 150 Science 1-Street, Urumqi 830011, China}

%% Note that the \and command from previous versions of AASTeX is now
%% depreciated in this version as it is no longer necessary. AASTeX 
%% automatically takes care of all commas and "and"s between authors names.

%% AASTeX 6.31 has the new \collaboration and \nocollaboration commands to
%% provide the collaboration status of a group of authors. These commands 
%% can be used either before or after the list of corresponding authors. The
%% argument for \collaboration is the collaboration identifier. Authors are
%% encouraged to surround collaboration identifiers with ()s. The 
%% \nocollaboration command takes no argument and exists to indicate that
%% the nearby authors are not part of surrounding collaborations.

%% Mark off the abstract in the ``abstract'' environment. 

\begin{abstract}
Type-C quasi-periodic oscillations (QPOs) in black hole X-ray transients typically manifest in the low-hard and hard-intermediate states. This study presents a detailed spectral and temporal analysis of the black hole candidate Swift J1727.8-1613 using NICER observations from August and September 2023, with a focus on the first flare period. {We detected Type-C QPOs whose centroid frequency increased from 0.33 Hz to 2.63 Hz. An additional increase in frequency was observed when the outburst entered a flare period.} The time-averaged spectra, along with the rms and phase-lag spectra of the QPOs, were jointly fitted using the time-dependent Comptonization model \texttt{vkompthdk} to examine the geometry of the corona during this flare. Correlations between spectral and temporal properties suggest that the detected type-C QPOs are primarily modulated by Lense-Thirring precession. {Leveraging simultaneous radio observations that indicate discrete jet ejections, we proposed a scenario to describe the co-evolution of the disk-corona-jet during a flare ($\sim$ 3 days). This scenario is partially supported for the first time by polarization data in the soft gamma-ray band from INTEGRAL/IBIS.} A phenomenological analysis of the corona scenario was also conducted.
\end{abstract}

%% Keywords should appear after the \end{abstract} command. 
%% The AAS Journals now uses Unified Astronomy Thesaurus concepts:
%% https://astrothesaurus.org
%% You will be asked to selected these concepts during the submission process
%% but this old "keyword" functionality is maintained in case authors want
%% to include these concepts in their preprints.

%% From the front matter, we move on to the body of the paper.
%% Sections are demarcated by \section and \subsection, respectively.
%% Observe the use of the LaTeX \label
%% command after the \subsection to give a symbolic KEY to the
%% subsection for cross-referencing in a \ref command.
%% You can use LaTeX's \ref and \label commands to keep track of
%% cross-references to sections, equations, tables, and figures.
%% That way, if you change the order of any elements, LaTeX will
%% automatically renumber them.
%%
%% We recommend that authors also use the natbib \citep
%% and \citet commands to identify citations.  The citations are
%% tied to the reference list via symbolic KEYs. The KEY corresponds
%% to the KEY in the \bibitem in the reference list below. 

\section{Introduction} \label{sec:intro}
{Black hole X-ray binaries (BHXBs) are compact systems consisting of a stellar-mass black hole accreting matter from a companion star via Roche lobe overflow. The innermost accretion flow is characterized by a geometrically thin, optically thick accretion disk and a hot, relativistic corona that serves as the Comptonizing medium \citep{Gilfanov2010}. As material spirals inward, viscous dissipation of gravitational energy heats the disk, producing multi-temperature blackbody emission predominantly observed in the soft X-ray band. A subset of these soft photons is Comptonized into harder X-rays in the corona, generating a non-thermal power-law spectrum that dominates at higher energies \citep[see][for a review]{Done2007A&ARv}. Additionally, a fraction of the Comptonized photons irradiates the inner accretion disk, where relativistic reprocessing generates reflection features \citep{Basko1974A&A, George1991MNRAS}. Recently, the X-ray reflection spectrum has become an important tool that served as a key diagnostic tool of spacetime curvature, as its sensitivity to relativistic effects enables precise tests of general relativity within a few gravitational radii of the black hole \citep{Bambi2021SSRv}.}

{BHXBs typically reside in quiescence for months to decades prior to outbursts, exhibiting X-ray luminosities orders of magnitude lower than their active phases. During outbursts, these systems trace an anticlockwise "q"-shaped trajectory in the hardness-intensity diagram (HID) \citep{Homan2001ApJS, Fender2004MNRAS}, marked by a hysteresis effect between spectral states. Four distinct accretion states are observationally defined through spectral-timing properties \citep{Homan2005Ap&SS}: the low-hard state (LHS), hard-intermediate state (HIMS), soft-intermediate state (SIMS), and high-soft state (HSS). A subset of systems may persist in anomalous states near peak luminosity \citep[e.g.,][]{Mendez1997ApJ, Belloni2005A&A, Motta2012MNRAS}. 
As the outburst starts, the source enters the LHS, dominated by a nonthermal power-law component and a relatively weak thermal disk component in the X-ray spectrum. As the accretion rate increases, the source becomes brighter and rapidly transitions through the HIMS and SIMS, eventually reaching the HSS. In HSS, the multi-temperature blackbody component dominates, while the hard component becomes steeper and weaker \citep{Remillard2006ARA&A}. Additionally, some sources may remain in the LHS and HIMS without ever entering the SIMS, leading to a so-called failed-transition outburst \citep{Alabarta2021MNRAS}. Finally, the source returns to the LHS and then back to quiescence as the accretion rate decreases.}

Relativistic jets can appear during these transitions between hard and soft states \citep[see e.g.][for a review of jets and X-ray binary outbursts]{Fender2004MNRAS, Remillard2006ARA&A}. Typically, two types of relativistic jets are observed in BHXBs, classified by their radio spectral index and morphology: a small-scale, optically thick, steady jet and an extended, optically thin, transient jet \citep[for a review]{Fender2006csxs.book}. A steady jet is usually present in the hard state, and even in the HIMS \citep{Fender2001MNRAS, Russell2019ApJ}. However, the steady jet emission is quenched as the system transitions from HIMS to SIMS \citep{Fender2004MNRAS, Russell2011ApJ}. During the transition to the soft state, a transient jet is launched due to discrete relativistic ejecta \citep{Mirabel1994Nature, Corbel2004ApJ, Miller-Jones2012MNRAS, Russell2019ApJ}.

In addition to their long-term evolution, BHXBs exhibit strong variability on sub-second timescales, which can be analyzed using Fast Fourier Transformation (FFT). A prominent feature of these systems is the presence of low-frequency quasi-periodic oscillations (LFQPOs; the centroid frequency is below 30 Hz) in the power density spectrum \citep[PDS; see e.g.,][for a review]{Ingram2019NewAR}. 
The main types of LFQPOs are categorized into types A, B, or C based on the shape and strength of the noise component in the PDS, as well as their root mean square (rms) amplitude and phase lags \citep{Wijnands1999ApJ, Remillard2002ApJ, Casella2005ApJ}. Type-C QPOs are the most common and strongest QPOs observed in BHXBs and typically appearing in the LHS and HIMS. They have rms amplitudes that can reach up to 20 percent and high-quality factors, with $Q \geq 10$ ($Q = \nu/\text{FWHM}$, where $\nu$ is the centroid frequency and FWHM is the full width at half-maximum of the QPO). Type-C QPOs frequently exhibit subharmonics, second and occasionally third harmonics, and are associated with a strong broad-band noise component in the PDS. The most popular model explaining these QPOs is the Lense–Thirring (LT) model proposed by \citet{Stella1998ApJ}, which attributes the QPO to a geometric effect under general relativity. 
% Alternative models suggest that type-C QPOs result from accretion-ejection instability in a magnetized disk \citep{Tagger1999A&A}, oscillations in a transition layer within the accretion flow \citep{Titarchuk2004ApJ}, or oscillations in the corona driven by magnetoacoustic waves \citep{Cabanac2010MNRAS}.

Phase lags provide additional insights into the X-ray variability in BHXBs. These lags are measured using the Fourier cross-spectrum, which is computed from light curves in two different energy bands \citep{Miyamoto1989Nature, Cui1997ApJ, Nowak1999ApJ}. Hard (positive) lags can occur due to the propagation of mass accretion rate fluctuations from the outer parts of the disk toward the inner disk and corona \citep[e.g.][]{Arevalo2006MNRAS, Ingram2013MNRAS}. Conversely, soft (negative) lags may be produced when the accretion disc is irradiated by hard photons from the corona, resulting in these photons being reprocessed and re-emitted by the disk at a later time compared to those directly observed from the corona. \citep[e.g.][]{Uttley2014A&ARv, Karpouzas2020MNRAS}. Thus, the LFQPOs can be one of the potential tools to shed light on the morphology of the inner accretion flow and its evolution.
\begin{deluxetable*}{cccc}
\tablenum{1}
\tablecaption{NICER observations of Swift J1727.8-1613 analyzed in present work. All observations were taken in 2023.\label{tab:observations}}
\tablewidth{0pt}
\tablehead{
\colhead{Obs} & \colhead{OBSID} & \colhead{Observation Time} & \colhead{Exposure (ks)}
}
% \decimalcolnumbers
\startdata
1 & 6203980102 & 2023-08-26 09:20:05.00 & 1.1 \\
2 & 6203980106 & 2023-08-30 00:00:20.00 & 10.6\\
3 & 6203980113 & 2023-09-06 10:05:00.00 & 5.4\\
4 & 6750010502 & 2023-09-08 00:40:24.00 & 4.8\\
5 & 6703010106 & 2023-09-12 00:24:50.00 & 0.8\\
6 & 6203980118 & 2023-09-12 09:43:16.00 & 2.0\\
7 & 6511080101 & 2023-09-13 01:12:37.00 & 1.1\\
8 & 6203980119 & 2023-09-13 04:18:37.00 & 4.9\\
9 & 6203980120 & 2023-09-14 00:26:33.00 & 6.3\\
10 & 6750010202 & 2023-09-15 07:30:40.00 & 6.5\\
11 & 6750010203 & 2023-09-16 03:34:15.00 & 1.8\\
\enddata
% \tablecomments{}
\end{deluxetable*}

Although the corona around a black hole is widely believed to consist of hot electrons with temperatures up to approximately 100 keV, these electrons give rise to the Comptonized spectrum \citep{Zdziarski1996MNRAS, zycki1999MNRAS}. However, many open questions remain regarding the evolution of the disk-corona-jet system. For instance, the process of disk truncation during the evolution of black hole transients is not fully understood \citep{Esin1997ApJ}, nor is the nature of the corona \citep{Galeev1979ApJ, Haardt1991ApJ, Markoff2005ApJ}. Understanding the geometry of the corona and its connection with the disk and the jet is vital to determining the physical nature of the Comptonization region. Evidence suggests a connection between the corona and the radio jet, as indicated by the universal radio–X-ray correlation in the LHS \citep{Gallo2003MNRAS, Fender2004MNRAS}. A study of GRS 1915+105 proposed that variations in the radio–X-ray correlation could be due to changes in the corona temperature \citep{Mendez2022NatAs}. \citet{You2021Nature} proposed a jet-like corona model for MAXI J1820+070 based on spectral analysis of the reflection component, interpreting the corona as a standing shock through which material flows.
Studies on X-ray variability have shown that the size of the corona evolves continuously during the outburst. These changes potentially link to variations in radio jet emission suggesting a disk-corona-jet coupling. \citep{Garcia2021MNRAS, Ma2021NatAs, Garcia2022MNRAS, Mendez2022NatAs, Zhang2022MNRAS, Fu2022RAA, Peirano2023MNRAS, Zhang2023MNRAS, ma2023variable, Yang2023MNRAS}.

\subsection{Swift J1727.8-1613}
The X-ray transient Swift J1727.8-1613, initially identified as GRB 230824A, was discovered by Swift/BAT on August 24, 2023 \citep{Page2023GCN}. A rapid increase in flux confirmed it as a new galactic X-ray transient \citep{2023ATel16205....1N, Nakajima2023ATel}. Observations across multiple wavelengths, including optical \citep{Castro-Tirado2023ATel}, X-ray \citep{OConnor2023ATel}, and radio \citep{Miller-Jones2023ATel}, suggest that Swift J1727.8-1613 is a low-mass black hole candidate. Further optical observations revealed that Swift J1727.8-1613 consists of a black hole primary with an early K-type companion star. It has an orbital period of approximately 7.6 hours and is located at a distance of $2.7 \pm 0.3$ kpc \citep{Mata2024A&A}. Additionally, IXPE detected polarized emission in the hard intermediate state of the source, with a polarization degree of $4.1\% \pm 0.2\%$ and a polarization angle of $2.2^\circ \pm 1.3^\circ$ \citep{Veledina2023ApJ}. Based on the polarization results, the inclination angle and the distance are predicted as $i \sim 30^\circ$-$60^\circ$ and 1.5 kpc, respectively. Additionally, the X-ray polarization angle aligns with observations in the submillimeter band, which suggests the Compton corona is elongated in the orthogonal direction of the jet emission \citep{Veledina2023ApJ, 2024ApJIngram}. A strong continuous jet has been observed in this source from polarization and radio analysis, which indicated the largest continuous jet observed in any X-ray binary \citep{2024ApJWood}.
Based on the spectral analysis from simultaneous observations with Insight-HXMT, NICER, and NuSTAR, \citet{Peng2024ApJL} estimated a black hole spin of approximately 0.98 and an orbital inclination of around 40 degrees. Temporal and spectral analyses by \citet{Yu2024MNRAS} and \citet{Chatterjee2024arXiv} suggested a high-inclination disk for this source. 

In this paper, we conducted temporal and spectral analyses of NICER data to investigate the evolution of the coronal geometry of Swift J1727.8-1613 during a flare state and examined the correlation between its temporal and spectral properties. Section \ref{sec:observation} describes the observations and data reduction procedures. Our temporal and spectral analysis are presented in Section \ref{sec:results}, followed by results and discussion in Section \ref{sec:discussion}. A summary is provided in Section \ref{sec:summary}.

\section{Observation and data reduction}\label{sec:observation}

The \textit{Neutron Star Interior Composition Explorer} (NICER) is a soft X-ray telescope onboard the International Space Station (ISS; \citealt{gendreau2016neutron}). NICER’s \textit{X-ray Timing Instrument} (XTI) covers the 0.2–12 keV energy band with an absolute timing precision of approximately 100 ns, making it an ideal instrument for studying fast X-ray variability. The XTI of NICER comprises an array of 56 co-aligned concentrator X-ray optics, each paired with a single-pixel silicon drift detector. Currently, 52 detectors are operational, with a peak effective area of approximately 1900 cm$^2$ at 1.5 keV.

Swift J1727.8-1613 has been observed with NICER almost daily since its discovery. In this work, we use 11 observations from NICER (see Table~\ref{tab:observations}). We processed the data using the NICER data analysis software (\texttt{NICERDAS v12a}) available in \texttt{HEASOFT V6.33.2}. This version includes updates addressing the optical light leak of May 2023. The task \texttt{nicerl2}\footnote{\url{https://heasarc.gsfc.nasa.gov/lheasoft/ftools/headas/nicerl2.html}} is used to generate clean event files, applying all standard calibration and data screening criteria. For some intervals of observations in the flare state, we found that the source flux changed significantly. 
To ensure stable temporal and spectral features from the flare state, we divided each observation into segments. We then selected intervals with exposure times $>$ 400 s, ensuring a relatively constant source count rate to study the temporal and spectral properties.
The intervals of each observation are given in Table \ref{tab:observations1}. Further, the spectral products are extracted using the \texttt{nicerl3-spect} tool. The background model \texttt{3c50} is selected using the flag \texttt{bkgmodeltype=3c50} during the extraction of spectral products. The detector redistribution matrix file and the auxiliary response file are generated with the tasks \texttt{nicerrmf} and \texttt{nicerarf}, respectively. The spectra are fitted in the 1–10 keV band using \texttt{XSPEC v12.14.0} because NICER data below 1 keV have significant residuals due to calibration issues. The timing properties are analyzed using the \texttt{stingray}\footnote{\url{https://github.com/StingraySoftware}} libraries \citep{Huppenkothen2019stingraya, Huppenkothen2019stingrayb, Bachetti2023stingray}.

\begin{figure}[ht!]
  \centering
  \begin{minipage}{0.49\linewidth}
    \centering
    \includegraphics[width=\linewidth]{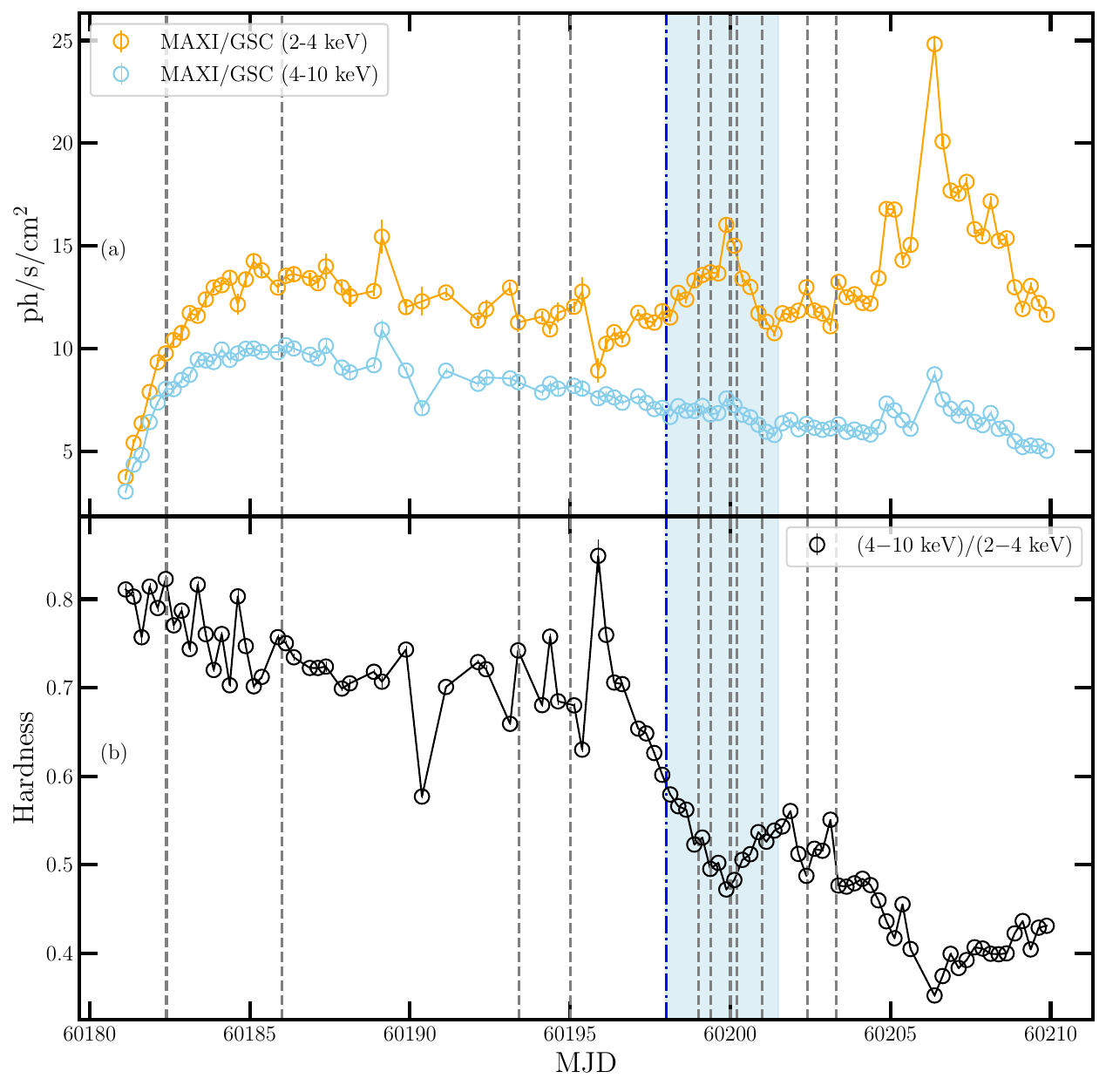}
  \end{minipage}
  \hfill
  \begin{minipage}{0.49\linewidth}
    \centering
    \includegraphics[width=\linewidth]{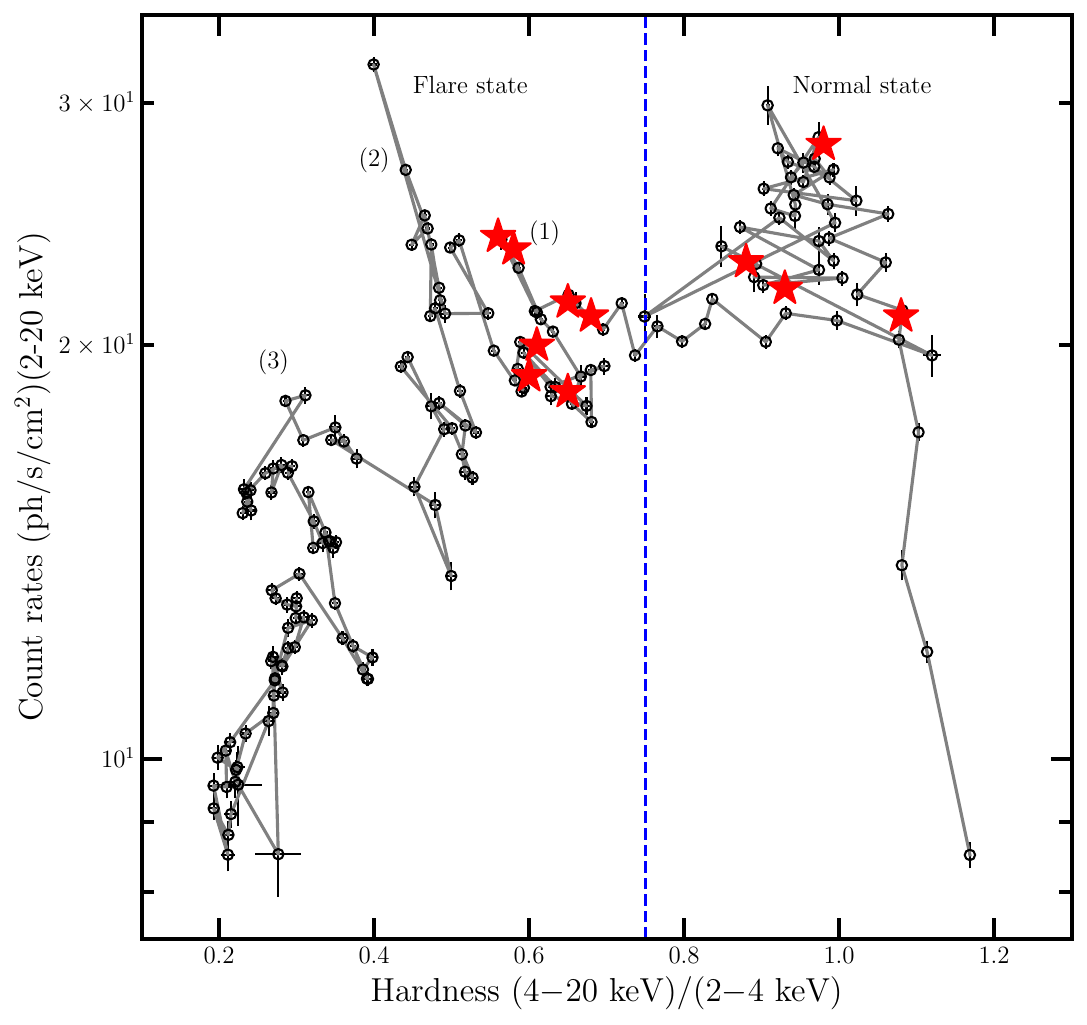}
  \end{minipage}
      \caption{{The left panel shows the variation of the (a) MAXI/GSC 2-4 keV count rate (orange) and MAXI/GSC 4-10 keV count rates (blue), and (b) hardness ratio with time. 
      The HR is the ratio of the 4-10 keV count rate to the 2-4 keV count rate of the MAXI/GSC data. The blue shadow area represents the first flare. The grey dashed lines connect the data points in time sequence. The right panel shows the HID of Swift J1727.8–1613 in this outburst observed by MAXI/GSC. Intensity is in the 2-20 keV. The hardness is defined as the ratio between 4–20 keV and 2–4 keV count rate. The observations analyzed in this work are marked with red stars.}}
  \label{fig:outburst_hardness}
\end{figure}

% \begin{figure}
%     \centering
% 	\includegraphics[width=\columnwidth]{images/lc_hardness.pdf}
%     \caption{Variation of the (a) MAXI/GSC 2-4 keV count rate, (b) MAXI/GSC 2-10 keV count rates, and (c) hardness ratio with time. The HR is the ratio of the 4-10 keV count rate to the 2-4 keV count rate of the MAXI/GSC data. The blue shadow represent the first flare.}
%     \label{fig:outburst}
% \end{figure}

% \begin{figure}
% 	\includegraphics[width=\columnwidth]{images/H_I.pdf}
%     \caption{The HID of Swift J1727.8–1613 in this outburst observed by MAXI/GSC. Intensity is in the 2-10 keV. The hardness is defined as the ratio between 4–10 keV and 2–4 keV count rate.}
%     \label{fig:H_I}
% \end{figure}

% Fig. ~\ref{fig:outburst}

\section{Data Analysis}\label{sec:results}

In this section, we perform timing and spectral analysis of the outburst of Swift J1727.8-1613 during 2023. We studied the QPO properties of the source during the outburst as well as its spectral nature and radiation properties using NICER data.

\subsection{The Evolution of Swift J1727.8-1613}

First, we extracted the daily average light curve using the MAXI/GSC3\footnote{\url{http://maxi.riken.jp/pubdata/v7.7l/J1727-162/index.html}} public archive. In the left panel of Fig.~\ref{fig:outburst_hardness}, we show the variation of the flux during the outburst. The outburst started roughly around MJD 60180 (2023 August 24), when the flux emerged from quiescence. The MAXI/GSC flux began to rise after this date, as seen in sub-panels (a) and (b) in the left panel of Fig.~\ref{fig:outburst_hardness}. Within five days, the flux increased very rapidly, reaching its peak on MJD 60185 (2023 August 29). Subsequently, the flux began to decrease very slowly, except for the period from MJD 60198 to 60210, where the flux showed significant variation. However, the variation in the 4-10 keV band is much smaller than in the 2-4 keV band, suggesting its origin from the soft component. In sub-panel (b), we show the variation of the hardness ratio (HR), which is the ratio of the hard 4-10 keV flux to the soft 2-4 keV flux. 
Following the approach of \citet{Yu2024MNRAS}, we divided the light curve into two states, the normal states and the flare states\footnote{{As in eruptions observed in other typical BHXBs, the initial state is defined as the 'normal state,' which exhibits a segment of the standard q-shaped pattern on the hardness-intensity diagram.} The transition from the normal state to the second state occurred around MJD 60198, as indicated by the blue dashed line in Fig.~\ref{fig:outburst_hardness}. This second state is marked by multiple flares and a gradual decrease in hardness, and is therefore referred to as the "flare state."}, before and after MJD 60198, respectively. 
{At the onset of the outburst, the HR was approximately 0.8, later it gradually decreased until MJD 60190.}
{After a brief increase, the HR dropped sharply near MJD 60195 and then stabilized at a constant level until the subsequent flare. This decrease in HR signifies a weakening of the hard components as the source transitioned into the flare state.}

The HID is shown in the right panel of Fig.~\ref{fig:outburst_hardness}. The outburst begins in the right part of the plot. As the intensity increases, the source on the HID moves towards the upper left. For clarity, we focus observations on the first flare in this work, which is marked in the right panel of Fig.~\ref{fig:outburst_hardness}.

\subsection{Timing analysis}\label{sec:timing}
% \subsubsection{Power density spectrum}\label{sec:psd}
We extracted the PDS using archival NICER/XTI data (see e.g. Table~\ref{tab:observations}) and processed them with \texttt{HENDRICS}\footnote{\url{https://hendrics.stingray.science/en/latest/}}, which allows for dead time correction by comparing the two modules \citep{Bachetti2015ApJ}. For generating the PDS, we utilized a time resolution of 1/256 s over 16 s intervals in the 1--10 keV energy range. This setup produces spectra up to the Nyquist frequency (128 Hz), which we then averaged across all segments. Given our focus on LFQPOs below 30 Hz, a 1/256 s resolution is adequate. The PDS is normalized to fractional rms amplitude \citep{Belloni1990A&A}, the formula we utilized is $\text{rms}=\sqrt{P(S+B)/S}$ \citep{Bu2015ApJ}, where S and B stand for source and background count rates, respectively, and P is the power normalized according to \citet{Miyamoto1991ApJ}.  We do not consider the background rate when converting the PDS to rms units, as it is negligible compared to the source rate in our observations. We logarithmically rebin the PDS such that each bin size is 1.01 times larger than the previous bin. The 1--10 keV energy range is divided into five energy bands: 1.0--1.5 keV, 1.5--2.8 keV, 2.8--5.0 keV, 5.0--6.5 keV, and 6.5--10 keV. We produce a PDS for each of these energy bands.
% in the unit of (rms/mean)$^2$ Hz$^{-1}$
% We fit the PDS with four Lorentzians representing the QPO fundamental, the second harmonic, and two broadband noise components. An extra Lorentzian is needed if there is a QPO subharmonic. 

% The PDS is fitted with four or five Lorentzian components \citep{Belloni2002ApJ}, with two Lorentzian functions used to fit the fundamental QPO and the second harmonic, and two more to fit the broad-band noise. A power-law is then used to model the continuous Poisson noise at higher frequencies.

\begin{figure}[ht!]
  \centering
  \begin{minipage}{0.31\linewidth}
    \centering
    \includegraphics[width=\linewidth]{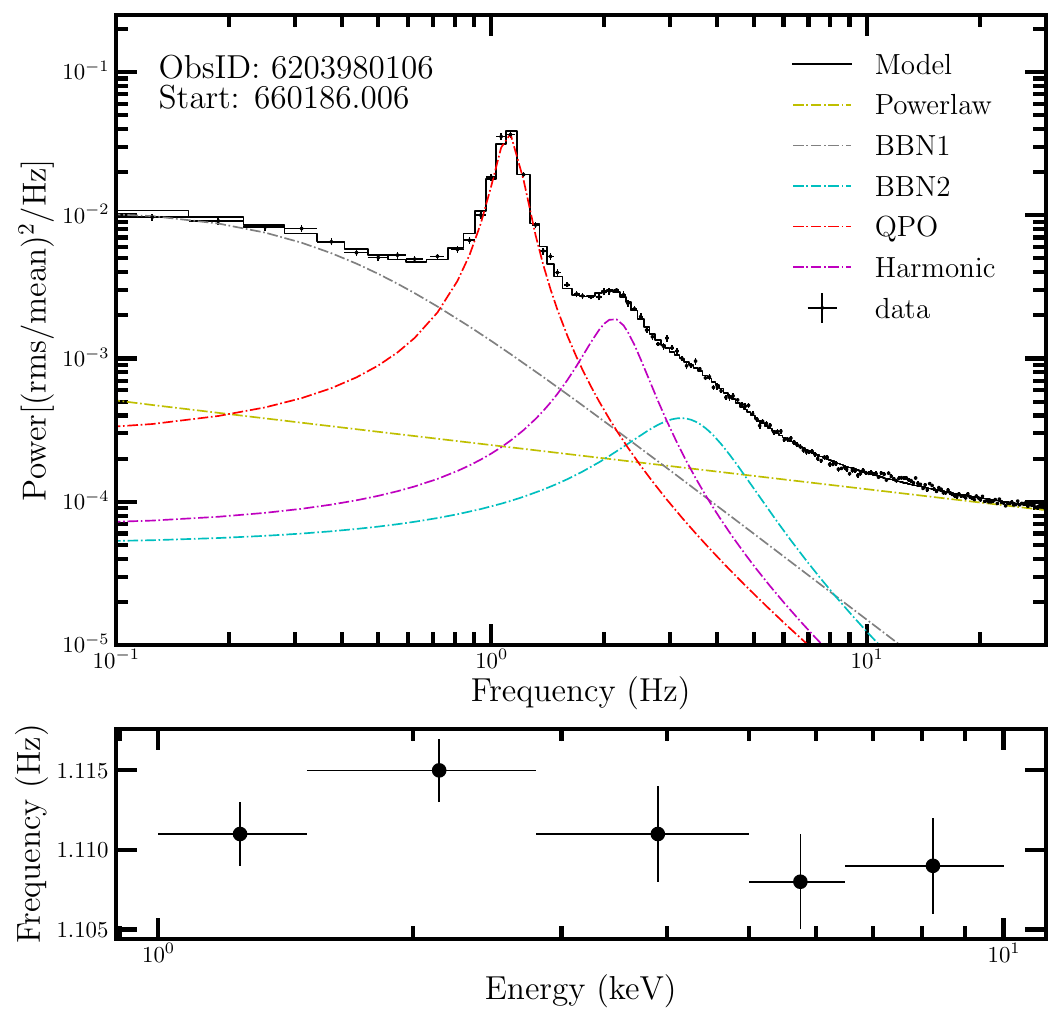}
  \end{minipage}
  \hfill
  \begin{minipage}{0.31\linewidth}
    \centering
    \includegraphics[width=\linewidth]{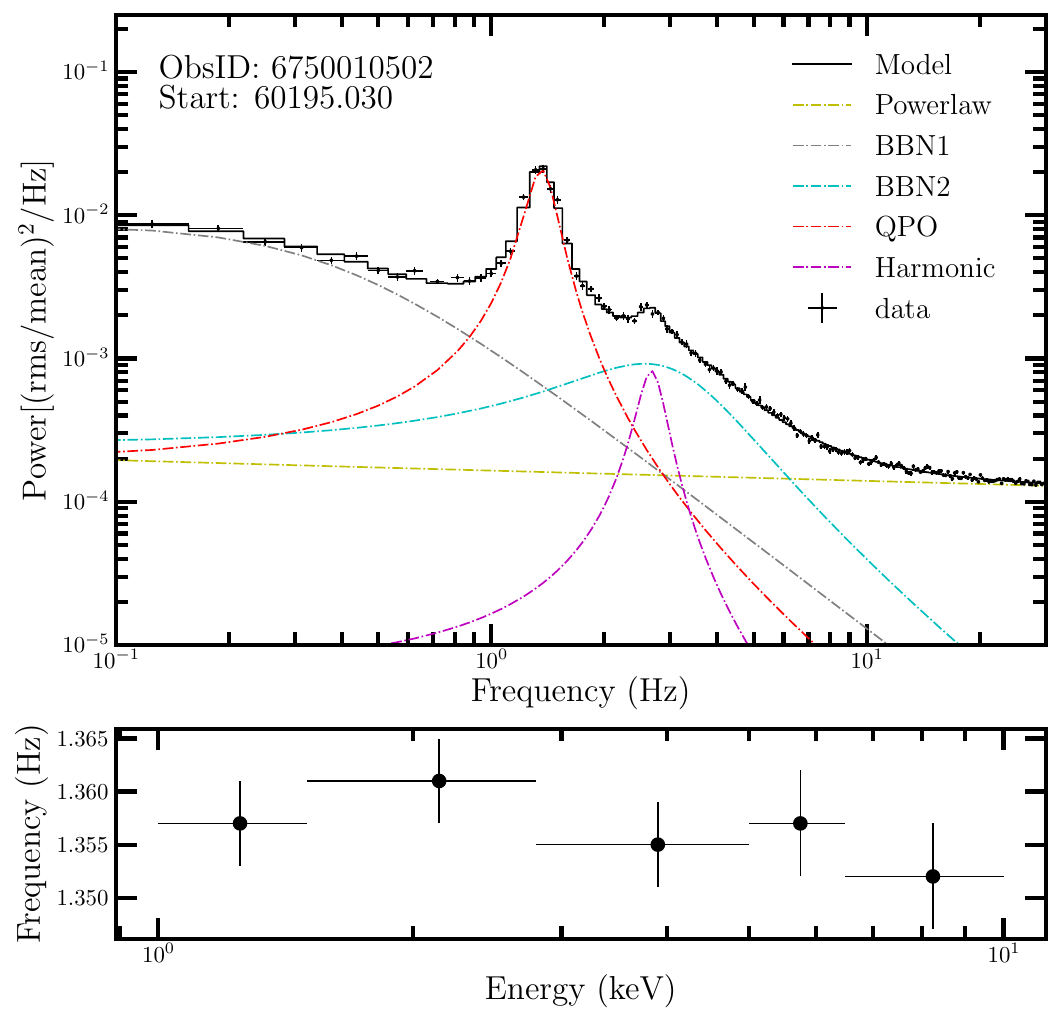}
  \end{minipage}
  \begin{minipage}{0.31\linewidth}
    \centering
    \includegraphics[width=\linewidth]{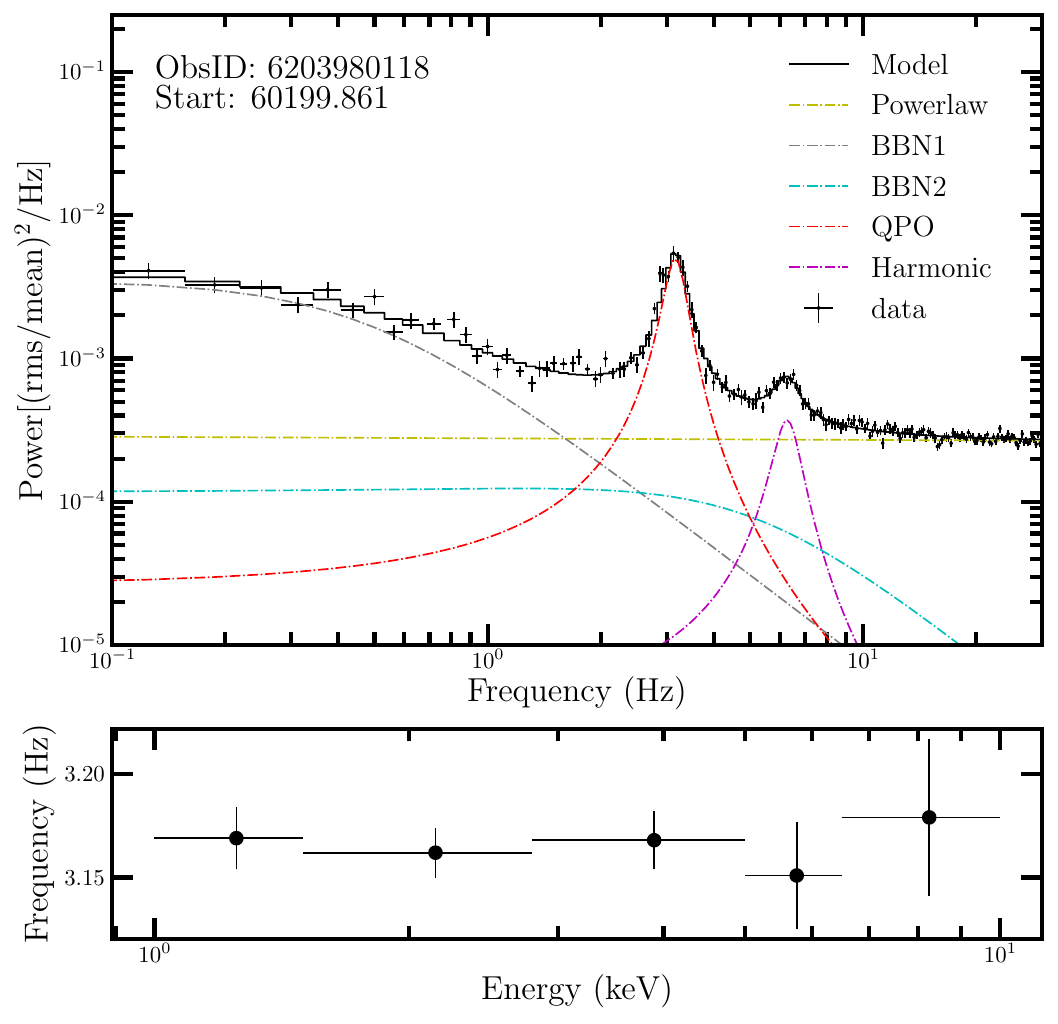}
  \end{minipage}
  
  \begin{minipage}{0.31\linewidth}
    \centering
    \includegraphics[width=\linewidth]{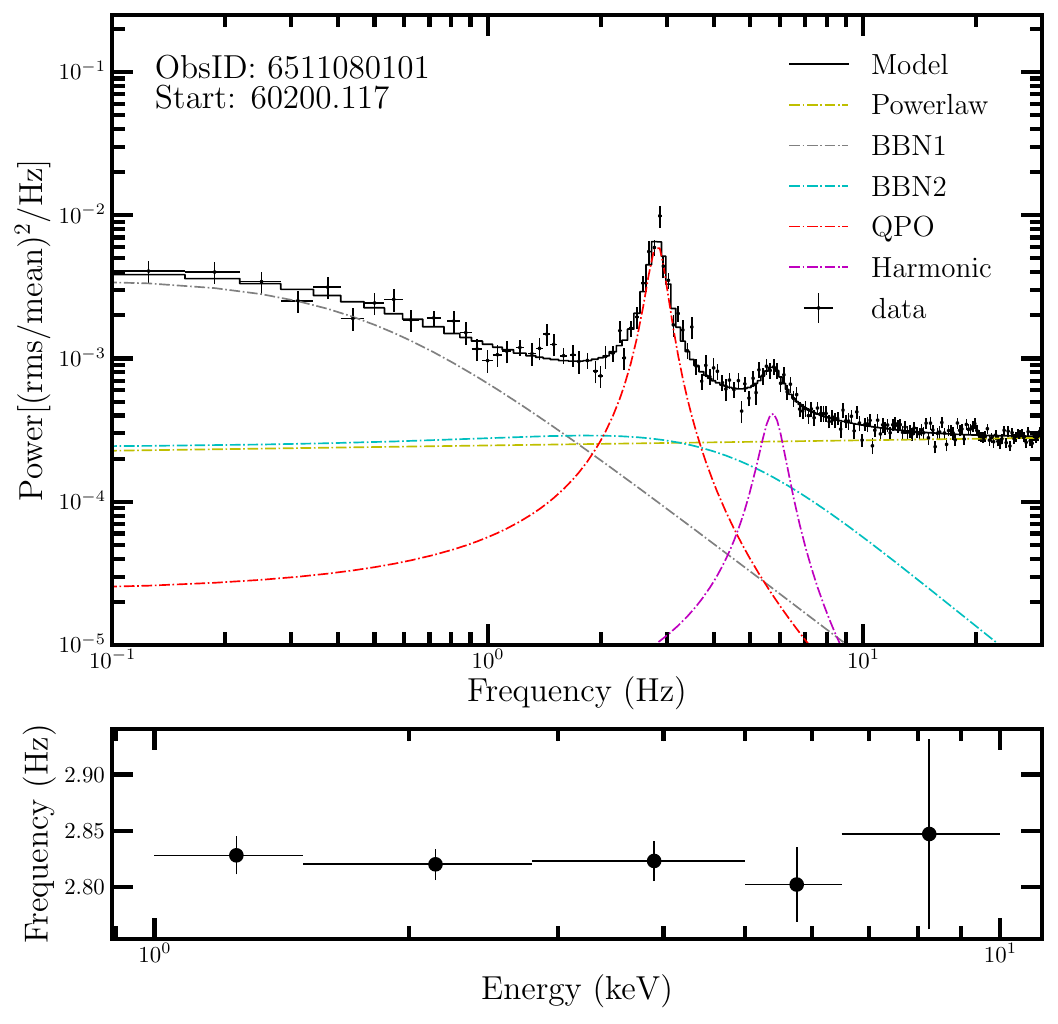}
  \end{minipage}
  \hfill
  \begin{minipage}{0.31\linewidth}
    \centering
    \includegraphics[width=\linewidth]{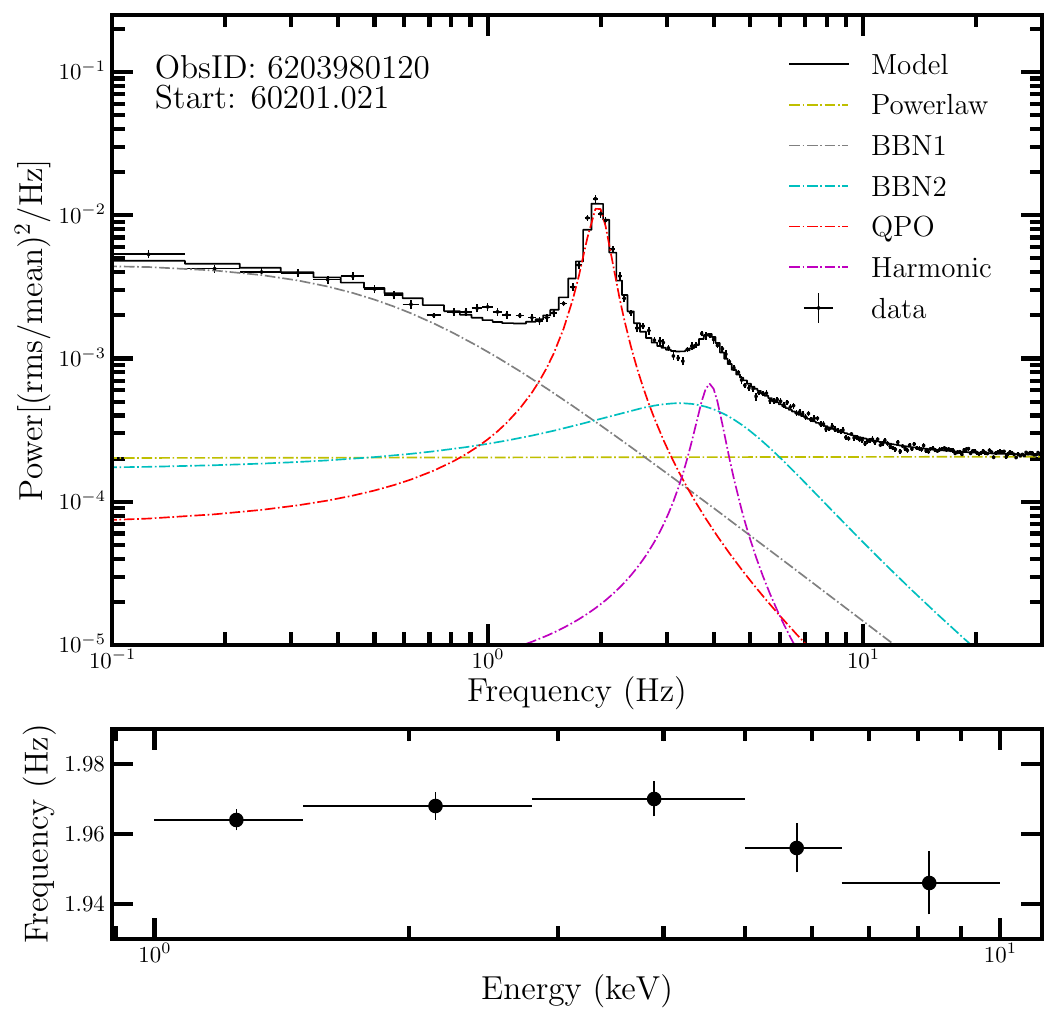}
  \end{minipage}
  \begin{minipage}{0.31\linewidth}
    \centering
    \includegraphics[width=\linewidth]{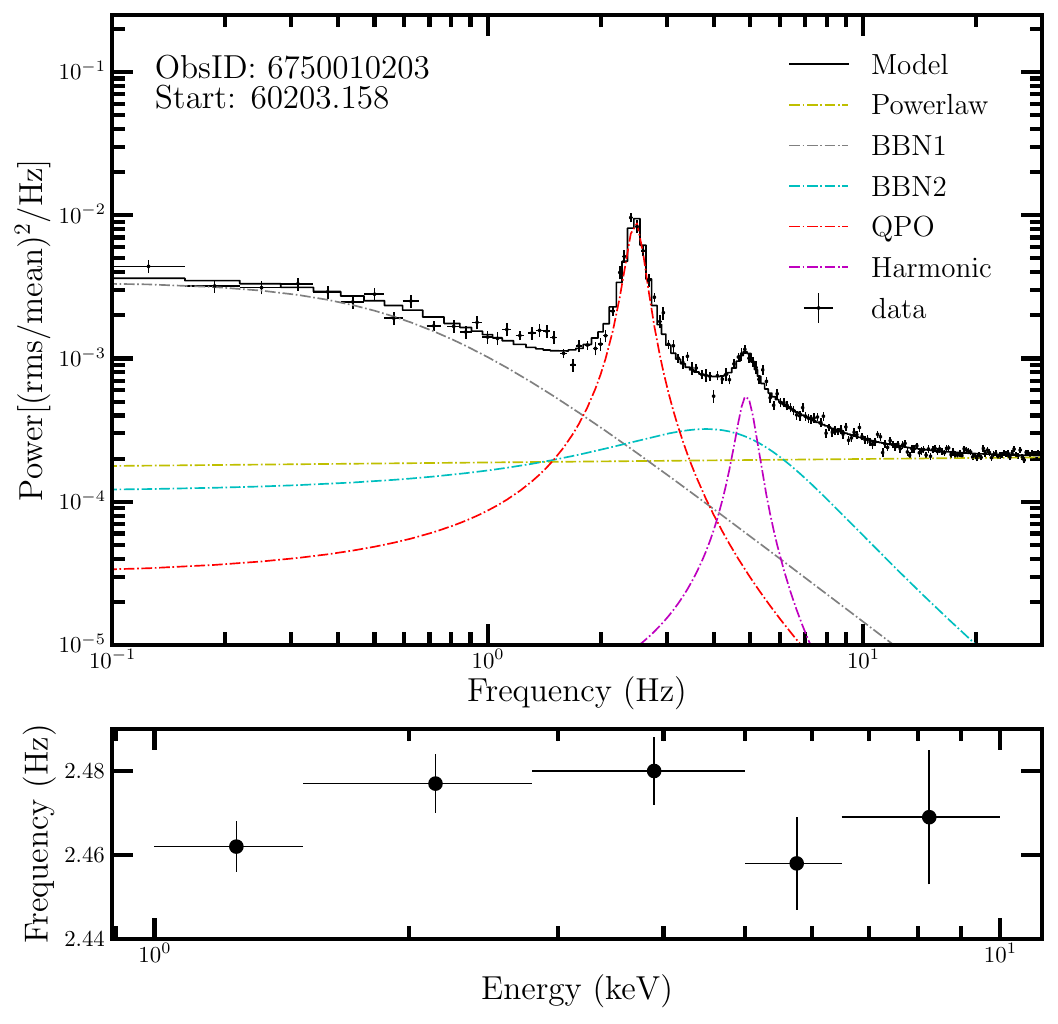}
  \end{minipage}
  
  \caption{{Six representative PDS of the outburst of Swift J1727.8–1613. The upper panel of each subplot shows the PDS in the 1–10 keV energy range, while the lower panel displays the fundamental QPO in each of the energy sub-bands: 1.0–1.5 keV, 1.5–2.8 keV, 2.8–5.0 keV, 5.0–6.5 keV, and 6.5–10 keV. Dashed lines represent the best-fitting Lorentzian functions and power-law component. BBN1 and BBN2 denote the two broadband noise components observed on different timescales.}}
  \label{fig:qpo}
\end{figure}

{The PDS is fitted with four or five Lorentzian components \citep{Belloni2002ApJ}. The fundamental QPO frequency and its second harmonic are fitted with two Lorentzian functions. The broad-band noise is also fitted with two more such components. A power-law is then used to model the continuous Poisson noise at higher frequencies.} 
We fit the power spectrum in XSPEC (details can be found in \citep{Ingram2012MNRAS}), and uncertainties are given for a 90 percent confidence interval. All parameters of a Lorentzian function (the central frequency, FWHM, and the normalization) are free, except for the central frequency of one Lorentzian component which fits the broadband noise at 0. 
% If no QPO is present, we reduce the number of Lorentzian components to one or two to fit only the broadband noise.
Using these models we fit the PDS of each orbit and then average them in the 1–10 keV band and in the six sub-bands (see above). 
% We fix the central frequencies and the FWHMs of all the Lorentzian components to the values obtained from the PDS in the 1–10 keV band and allow only the normalizations to be free when fitting the PDS in the narrower energy bands.
QPOs are identified by fitting a multi-Lorentzian model to the entire power spectrum and searching for features with a quality factor $Q = \nu / \delta\nu > 3$ and significance $\text{Norm} / \text{Err}_{\text{norm}} > 3$, where $\nu$ is the Lorentzian centroid frequency, $\delta\nu$ is FWHM, and $\text{Err}_{\text{norm}}$ is the positive error in the normalization.
We verify that the central frequency and width of the Lorentzian components do not change significantly with energy (see e.g., Fig.~\ref{fig:qpo}). We calculate rms amplitude of the variability components by taking the square root of the normalizations of the Lorentzians. The best-fit results for the PDSs are listed in Table \ref{tab:observations1}. Six representative power spectra, along with the energy dependence of their central QPO frequencies, are illustrated in Fig.~\ref{fig:qpo}.

\begin{deluxetable*}{ccccccccccccc}
\tablenum{2}
\tablecaption{Observation log and parameters of the QPOs of Swift J1727.8–1613. The error bar indicates the 90\% confidence level. All the quantities mentioned in the table have their usual meanings. See the text for details.\label{tab:observations1}}
\tablewidth{80pt}
\tabletypesize{\tiny}
% \tablewidth{0pt}
\tablehead{
\colhead{Obs} & \colhead{MJD start} & \colhead{MJD stop} & \colhead{$f^a$ (Hz)}  & \colhead{$\sigma^a$ (Hz)} & \colhead{rms$^a$ (\%)} & \colhead{$f^b$ (Hz)} & \colhead{$\sigma^b$ (Hz)} & \colhead{rms$^b$ (\%)} & \colhead{$\sigma_{\text{BBN1}}$ (Hz)}& \colhead{$f_{\text{BBN2}}$ (Hz)}& \colhead{$\sigma_{\text{BBN2}}$ (Hz)}& \colhead{$\chi^2$/dof}
}
% \decimalcolnumbers
\startdata
1  & 60182.587 & 60182.849& $0.33 \pm 0.01$ & $0.05 \pm 0.01$ & $9.0 \pm 2.9$ & $0.72 \pm 0.03$ & $0.28 \pm 0.09$& $4.9 \pm 2.8$ & $0.23 \pm 0.05$ & $4.75 \pm 0.19$ & $2.19 \pm 0.83$ & $253/195$\\
2  & 60186.006 & 60186.655 & $1.12 \pm 0.01$ & $0.19 \pm 0.01$ & $10.5 \pm 1.6$ & $2.11 \pm 0.01$ & $0.80 \pm 0.07$& $4.7 \pm 1.4$  & $0.80 \pm 0.06$ &$3.20 \pm 0.14$ & $2.47 \pm 0.15$& $247/171$ \\
3  & 60193.424 & 60193.946 & $1.42 \pm 0.01$ & $0.24 \pm 0.01$ & $9.2 \pm 1.3$ & $2.72 \pm 0.02$ & $0.95 \pm 0.10$& $4.5 \pm 1.2$  & $0.99 \pm 0.06$& $4.18 \pm 0.12$ & $2.62 \pm 0.18$ & $242/166$ \\
4 & 60195.030 & 60195.552 & $1.40 \pm 0.01$ & $0.26 \pm 0.01$ & $9.1 \pm 1.5$ & $2.67 \pm 0.02$ & $0.47 \pm 0.11$& $2.4 \pm 1.3$  & $0.80 \pm 0.06$ & $2.57\pm 0.11$ & $3.16\pm 0.17$ & $233/161$ \\
5  & 60199.217 & 60199.346 & $2.76 \pm 0.02$ & $0.27 \pm 0.05$ & $6.0 \pm 2.0$ & $5.58 \pm 0.13$ & $0.88 \pm 0.45$& $2.1 \pm 1.5$  & $1.21 \pm 0.40$ & $2.44 \pm 2.43^*$ & $11.83 \pm 5.05^*$ & $148/163$ \\
6  & 60199.861 & 60199.995 & $3.16 \pm 0.01$ & $0.46 \pm 0.03$ & $6.0 \pm 1.3$ & $6.27 \pm 0.05$ & $1.14 \pm 0.18$ & $2.5 \pm 1.0$  & $0.96 \pm 0.12$ & $0.66 \pm 14.74^{*}$ & $12.10 \pm 16.46^{*}$ & $182/163$ \\
7  & 60200.117 & 60200.124 & $2.83 \pm 0.01$ & $0.39 \pm 0.04$ & $6.1 \pm 1.5$ & $5.70 \pm 0.06$ & $1.23 \pm 0.22$& $2.9 \pm 1.1$  & $0.96 \pm 0.16$ & $1.87 \pm 1.78$ & $8.01 \pm 2.26$ & $216/208$ \\
8  & 60200.377 & 60200.512 & $2.61 \pm 0.01$ & $0.29 \pm 0.02$ & $6.3 \pm 1.2$ & $5.23 \pm 0.02$ & $0.71 \pm 0.08$& $2.4 \pm 0.7$  & $1.09 \pm 0.15$ & $1.89 \pm 0.68$& $6.65 \pm 1.58$ & $195/161$ \\
9  & 60201.021 & 60201.480 & $1.97 \pm 0.01$ & $0.27 \pm 0.01$ & $7.0 \pm 1.2$ & $3.92 \pm 0.02$ & $0.72 \pm 0.08$& $2.8 \pm 0.9$  & $1.24 \pm 0.07$ & $2.96 \pm 0.29$ & $ 5.44 \pm 0.51$ & $233/161$ \\
10  & 60202.315 & 60202.850 & $2.63 \pm 0.01$ & $0.52 \pm 0.02$ & $6.5 \pm 1.0$ & $5.17 \pm 0.03$ & $1.59 \pm 0.16$& $3.1 \pm 1.1$ & $1.19 \pm 0.06$ & $3.70 \pm 1.24$ & $11.69 \pm 2.47$ & $246/174$ \\
11  & 60203.158 & 60203.237 & $2.47 \pm 0.01$ & $0.27 \pm 0.02$ & $6.3 \pm 1.3$ & $4.89 \pm 0.03$ & $0.68 \pm 0.10$ & $2.4 \pm 0.9$ & $1.39 \pm 0.13$ & $3.37 \pm 0.74$& $7.21 \pm 1.20$ & $208/161$ \\
\enddata
\raggedright 
% \footnotemark[1] represents the QPO component.\\
% \footnotemark[2] represents the Harmonic component.\\
\tablecomments{The marks $^a$ and $^b$ indicate the fundamental QPO and Harmonic properties. The frequency for BBN1 is fixed at 0. $^{*}$ indicates a large error estimated from the BBN due to the data quality.}
\end{deluxetable*}

% \subsubsection{Fractional rms and Phase-lag spectra}
% As shown in Fig.~\ref{fig:qpo}, we fit the PDS with four Lorentzians representing two broadband noise components, the QPO, and the second harmonic. Table \ref{tab:observations1} gives the best-fitting parameters of the Lorentzians within all the observation data. The second harmonic of QPO appears at a central frequency, consistent with being twice the central frequency of the QPO fundamental. To study the rms and phase lag of LFQPO and its energy dependence, we calculated the rms and lag energy spectra for each energy sub-band (see e.g. Fig.~\ref{fig:rms_lag}). As the energy increases from 1 keV to 10 keV, the rms increases monotonically from below 7 percent to 13 percent, indicating that the type-C QPO is mainly modulated by the hot corona.

\begin{figure}[ht!]
  \centering
  \begin{minipage}{0.49\linewidth}
    \centering
    \includegraphics[width=\linewidth]{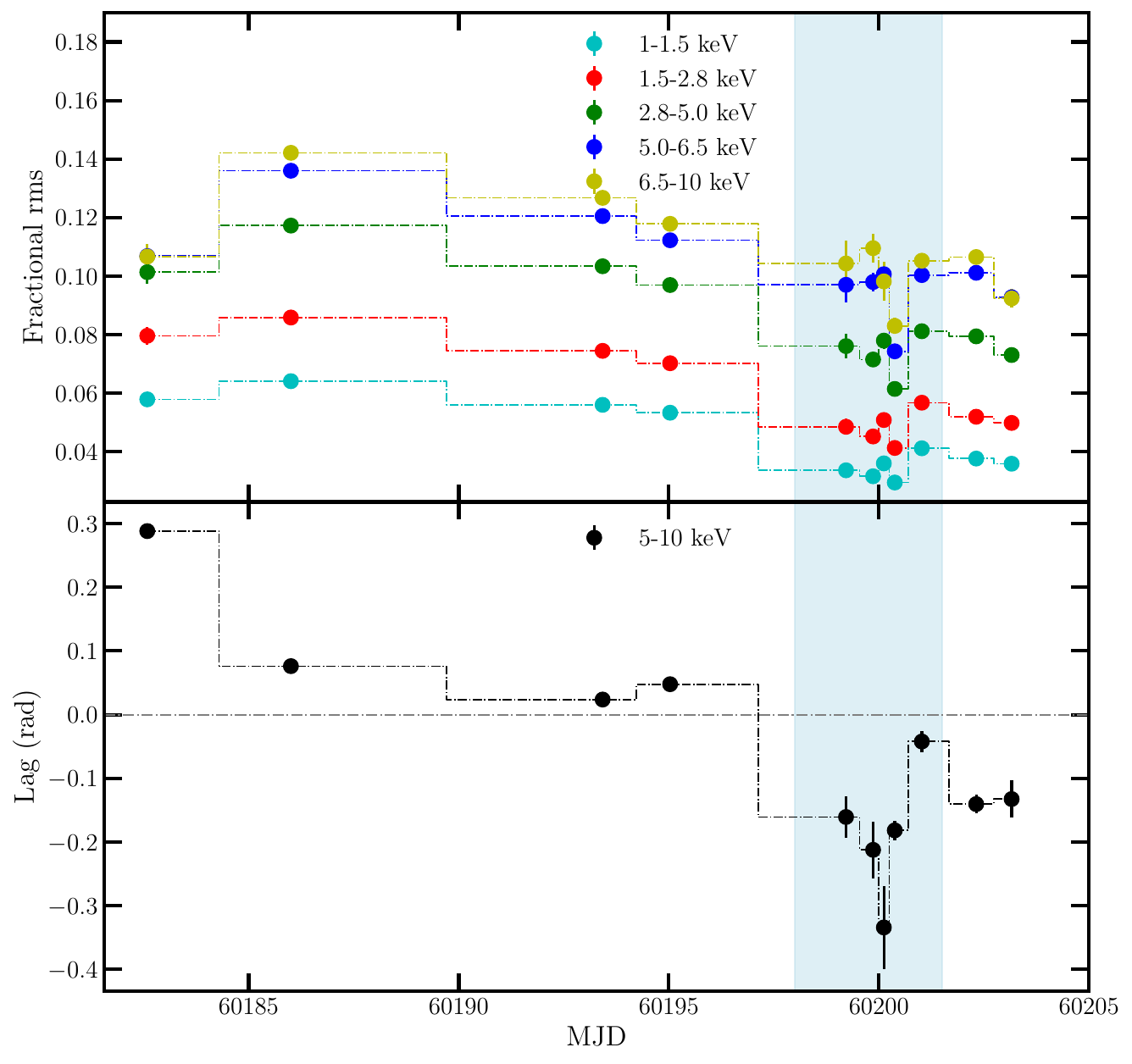}
  \end{minipage}
  \hfill
  \begin{minipage}{0.49\linewidth}
    \centering
    \includegraphics[width=\linewidth]{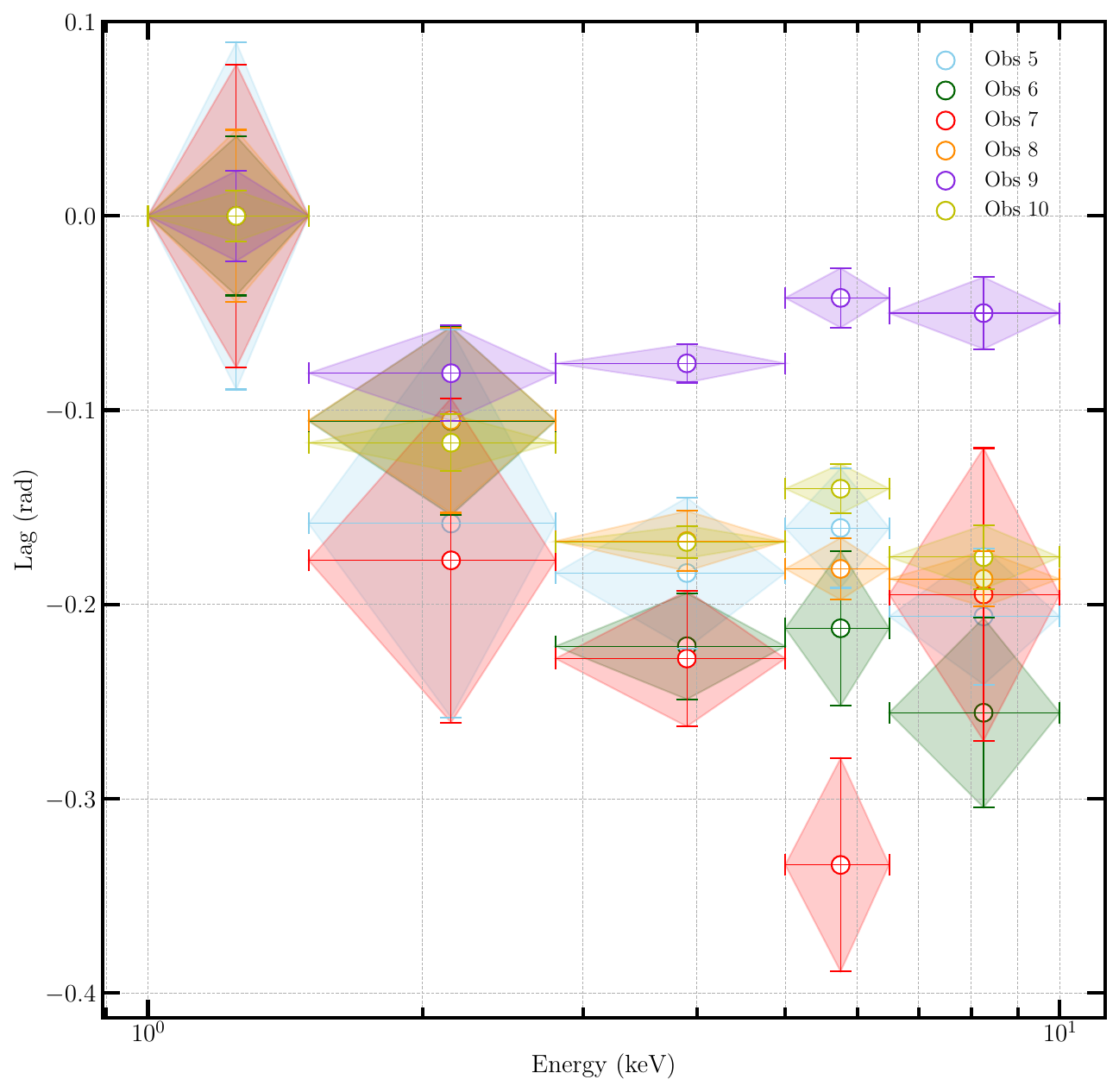}
  \end{minipage}

  \caption{{The left panel shows the evolution of the fractional rms amplitude and phase lags of the LFQPO from Swift J1727.8–1613. The blue-shaded region indicates the period of the first flare. The right panel displays the phase-lag energy spectrum of the LFQPO. The 1–1.5 keV band is used as the reference for calculating the phase lags.}}
  \label{fig:rms_lag}
\end{figure}

{Phase lag spectra for each observation were derived from NICER/XTI data, adopting identical time resolutions and segment lengths as those applied in the PDS calculations. The cross spectrum is defined as $CF(j) = f_1^*(j) \times f_2(j)$, where $ f_1(j) $ and $ f_2(j) $ represent the complex Fourier coefficients of the two energy bands at frequency $ \nu_j $, with $ f_1^*(j) $ denoting the complex conjugate of $ f_1(j) $ \citep{1987ApJvan}. Phase lags $ \phi_j = \arg[CF(j)] $ at each Fourier frequency $ \nu_j $ yield corresponding time lags $ \tau_j = \phi_j/(2\pi\nu_j) $ \citep{2014A&ARvUttley}. To quantify the energy-dependent phase lag relative to the reference band, we averaged $ \phi_j $ over the frequency interval $ [\nu_{\mathrm{QPO}} - \Delta\nu_{\mathrm{QPO}}/2, \, \nu_{\mathrm{QPO}} + \Delta\nu_{\mathrm{QPO}}/2] $, where $ \nu_{\mathrm{QPO}} $ and $ \Delta\nu_{\mathrm{QPO}} $ correspond to the centroid frequency and FWHM of the Lorentzian component of the PDS (see Fig.~\ref{fig:rms_lag} panels).}
A positive (hard) lag indicates that the hard photons lag behind the soft ones. The phase lag with respect to the 1–1.5 keV reference band crosses the zero line, as shown in the upper left corner of the right panel of Fig.~\ref{fig:rms_lag}.

\subsection{Time-averaged spectrum analysis and fitting with vKompthdk}
Next, we study the evolution of spectrum properties. To analyze the spectrum of Swift J1727.8–1613, we use the energy band of 1--1.7, 2.1--2.2, \& 2.3--10 keV to avoid the calibration in the Si band (1.7–2.1 keV) and the Au band (2.2–2.3 keV). We then fitted jointly the time-averaged energy spectrum, and the rms and phase-lag spectra of the type-C QPO, using the model \texttt{tbabs$\times$(diskbb+nthComp)+vkompthdk$\times$dilution}. \texttt{tbabs} describes the galactic absorption column density along the line of sight. \texttt{diskbb} is a multi-temperature blackbody component \citep{mitsuda1984energy}. \texttt{vkompthdk}\footnote{\url{https://github.com/candebellavita/vkompth}} is a single-component Comptonization model developed by \citet{Bellavita2022MNRAS}, similar to \texttt{nthcomp} \citep{Zdziarski1996MNRAS}. This model assumes that the source of the seed photons is a geometrically thin and optically thick accretion disk \citep{Shakura1973A&A}.
In this model, $kT_{s}$ represents the seed photon temperature of the Comptonized component. These seed photons undergo inverse-Compton scattering within a spherically symmetric corona, characterized by a size $L$ and an electron temperature $kT_e$. The corona is kept in thermal equilibrium by heating from an external source. A feedback process is also included, where a fraction of the photons scattered into the corona is redirected back onto the accretion disc. The feedback fraction, $\eta$, ranges from 0 to 1 and indicates the fraction of the disk flux resulting from this feedback \citep{Karpouzas2020MNRAS}. It is related to the intrinsic feedback fraction, $\eta_{\rm int}$, which represents the proportion of photons emitted by the corona that return to the accretion disk. The model treats the QPO as small oscillations in the spectrum around the time-averaged one. It assumes that spectrum fluctuations are caused by perturbations in the electron temperature, $kT_e$, through feedback mechanisms involving the seed photon temperature, $kT_s$, due to an oscillating external heating rate, $\delta\dot{H}_{ext}$ (for details, see \citealt{Bellavita2022MNRAS}).

{The \texttt{dilution} component is a correction to the fractional rms amplitude computed in the model to take into account the fraction of the non-variable emission. It is described by  Flux$_{\text{Compt}}$ / Flux$_{\text{Total}}$ such that RMS$_{\text{Obs}}$ = RMS$_{\text{Compt}}$ * Flux$_{\text{Compt}}$ / Flux$_{\text{Total}}$.}
Note that this \texttt{dilution} component does not introduce any new parameters into the fits. 
{The main parameters of the model are the seed photon temperature ($kT_s$), the electron temperature ($kT_e$), the power-law photon index ($\Gamma$), the size of the corona ($L$), the feedback fraction ($\eta$), and the variation of the external heating rate ($\delta\dot{H}_{ext}$).}

In our joint fitting analysis, we employed the external component \texttt{vkompthdk} to fit the rms and lag spectra of the QPO as well as the spectrum of Swift J1727.8--1613. We linked the seed photon temperature, $kT_s$, of \texttt{vkompthdk} to the inner disk temperature, $kT_{bb}$, of \texttt{diskbb}. Additionally, we linked the electron temperature and the power-law photon index of \texttt{vkompthdk} to the corresponding parameters of \texttt{nthcomp}. Given that the high-energy cutoff of the Comptonized component is beyond the NICER energy range, we fixed the electron temperature at 50 keV. The best-fitting parameters from all observations are presented in Table ~\ref{tab:spectrumfit}.

\begin{deluxetable}{p{1.8cm}p{1.cm}p{1.cm}p{1.cm}p{1.cm}p{1.cm}p{1.cm}p{1.cm}p{1.cm}p{1.cm}p{1.cm}p{1.cm}}
\tablenum{3}
\tablecaption{Fitting results of Swift J1727.8-1613.\label{tab:spectrumfit}}
\tablewidth{0pt}
\tabletypesize{\tiny}
\tablehead{
\colhead{Component} & \colhead{Obs1} & \colhead{Obs2} & \colhead{Obs3} & \colhead{Obs4} & \colhead{Obs5} & \colhead{Obs6} & \colhead{Obs7} & \colhead{Obs8} & \colhead{Obs9} & \colhead{Obs10} & \colhead{Obs11}
}
\startdata
\texttt{tbabs} & & & & & & & & & & & \\
$N_{\text{H}}$ ($\times 10^{21}\text{cm}^{-2}$) 
& $2.25^{+0.14}_{-0.14}$ & $2.70^{+0.11}_{-0.11}$ & $2.29^{+0.09}_{-0.09}$ 
& $2.31^{+0.09}_{-0.09}$ & $1.94^{+0.05}_{-0.04}$ & $1.87^{+0.02}_{-0.02}$ 
& $1.81^{+0.04}_{-0.03}$ & $1.79^{+0.02}_{-0.02}$ & $1.96^{+0.07}_{-0.08}$ 
& $1.94^{+0.07}_{-0.08}$ & $1.92^{+0.09}_{-0.07}$ \\
\hline
\texttt{diskbb} & & & & & & & & & & & \\
$k T_{\text{in}} (\text{keV})$ 
& $0.276^{+0.005}_{-0.004}$ & $0.319^{+0.005}_{-0.005}$ & $0.364^{+0.006}_{-0.006}$ 
& $0.360^{+0.006}_{-0.006}$ & $0.491^{+0.011}_{-0.012}$ & $0.525^{+0.008}_{-0.007}$ 
& $0.526^{+0.009}_{-0.009}$ & $0.480^{+0.005}_{-0.006}$ & $0.431^{+0.007}_{-0.006}$ 
& $0.450^{+0.010}_{-0.009}$ & $0.447^{+0.010}_{-0.012}$ \\
Norm ($\times 10^{2}$) 
& $4253^{+750}_{-604}$ & $3664^{+236}_{-415}$ & $1539^{+166}_{-150}$ 
& $1607^{+172}_{-154}$ & $252^{+12}_{-10}$ & $177^{+7}_{-7}$ 
& $255^{+8}_{-7}$ & $280^{+7}_{-6}$ & $567^{+43}_{-46}$ 
& $308^{+23}_{-24}$ & $368^{+37}_{-27}$ \\
\hline
\texttt{vkompthdk (nthcomp)} & & & & & & & & & & & \\
$\Gamma$ 
& $1.597^{+0.003}_{-0.004}$ & $1.793^{+0.002}_{-0.003}$ & $1.855^{+0.003}_{-0.003}$ 
& $1.846^{+0.003}_{-0.003}$ & $2.181^{+0.011}_{-0.012}$ & $2.237^{+0.007}_{-0.009}$ 
& $2.134^{+0.012}_{-0.013}$ & $2.124^{+0.005}_{-0.005}$ & $1.980^{+0.004}_{-0.003}$ 
& $2.179^{+0.004}_{-0.004}$ & $2.141^{+0.006}_{-0.006}$ \\
$L$ ($10^3$ km) 
& $-$ & $-$ & $-$ & $-$ & $1.58^{+0.92}_{-0.91}$ & $2.23^{+0.46}_{-0.45}$ 
& $2.71^{+0.90}_{-0.91}$ & $1.55^{+0.46}_{-0.45}$ & $< 0.95$ 
& $2.03^{+0.19}_{-0.19}$ & $2.09^{+0.62}_{-0.60}$ \\
$\eta$ 
& $-$ & $-$ & $-$ & $-$ & $0.41^{+0.04}_{-0.03}$ & $0.39^{+0.02}_{-0.01}$ 
& $0.48^{+0.03}_{-0.03}$ & $0.35^{+0.02}_{-0.02}$ & $0.49^{+0.01}_{-0.01}$ 
& $0.43^{+0.01}_{-0.01}$ & $0.48^{+0.02}_{-0.02}$ \\
$\delta\dot{H}_{ext}$ 
& $-$ & $-$ & $-$ & $-$ & $0.14^{+0.01}_{-0.01}$ & $0.14^{+0.01}_{-0.01}$ 
& $0.14^{+0.01}_{-0.01}$ & $0.14^{+0.01}_{-0.01}$ & $0.14^{+0.01}_{-0.01}$ 
& $0.14^{+0.01}_{-0.01}$ & $0.12^{+0.01}_{-0.01}$ \\
Norm 
& $22.7^{+0.2}_{-0.2}$ & $39.2^{+0.4}_{-0.4}$ & $33.9^{+0.4}_{-0.4}$ 
& $31.9^{+0.4}_{-0.4}$ & $40.4^{+1.4}_{-1.3}$ & $41.4^{+0.9}_{-1.0}$ 
& $34.8^{+1.0}_{-1.0}$ & $35.7^{+0.5}_{-0.5}$ & $28.7^{+0.5}_{-0.4}$ 
& $37.0^{+0.8}_{-0.9}$ & $34.5^{+1.0}_{-0.8}$ \\
\hline
$\chi^2$/$\text{dof}$ 
& $962/894$ & $860/840$ & $698/840$ & $677/840$ & $840/790$ 
& $1071/847$ & $874/803$ & $1080/787$ & $612/845$ 
& $591/845$ & $605/845$ \\
\hline
log$F_{\text{disk}}$ $\text{erg} \, \text{s}^{-1} \, \text{cm}^{-2}$
& $-8.036 \pm 0.026$ & $-7.714 \pm 0.017$ & $-7.760 \pm 0.014$ 
& $-7.772 \pm 0.013$ & $-7.850 \pm 0.052$ & $-7.852 \pm 0.042$ 
& $-7.695 \pm 0.033$ & $-7.855 \pm 0.022$ & $-7.797 \pm 0.012$ 
& $-7.957 \pm 0.031$ & $-7.895 \pm 0.031$ \\
log$F_{\text{corona}}$ $\text{erg} \, \text{s}^{-1} \, \text{cm}^{-2}$
& $-6.816 \pm 0.001$ & $-6.677 \pm 0.001$ & $-6.750 \pm 0.001$ 
& $-6.775 \pm 0.001$ & $-6.729 \pm 0.005$ & $-6.716 \pm 0.004$ 
& $-6.760 \pm 0.004$ & $-6.771 \pm 0.002$ & $-6.842 \pm 0.001$ 
& $-6.791 \pm 0.003$ & $-6.811 \pm 0.004$ \\
\enddata
\tablecomments{Best-fitting parameters of the joint fit to the energy spectrum of Swift J1727.8-1613, and the rms and lag spectra of the LFQPO in present work, using the single-component Comptonization model \texttt{vkompthdk}. The error indicates the 1-$\sigma$ confidence level. See the text for more details about the parameters.}
\end{deluxetable}

\section{Results and Discussion}\label{sec:discussion}
We performed a comprehensive spectral-timing analysis of Type-C QPOs observed during both the normal state and the first flare of Swift J1727.8-1613. This study provides the first comparison of the rms and lag spectra of low-frequency QPOs in a black hole X-ray binary during such a flare state. 
{By fitting the time-averaged energy spectra of the source, as well as the rms and lag spectra of the QPOs in each observation, we used the time-dependent Comptonization model \texttt{vkompth} \citep{Bellavita2022MNRAS} to infer the evolution of the physical parameters. For the first time, we reveal correlations between these parameters and the geometrical properties of the corona in a black hole candidate during a flare period.}
Finally, we propose a scenario for the evolution of the Comptonized region.

\begin{figure}
    \centering
	\includegraphics[width=0.49\linewidth]{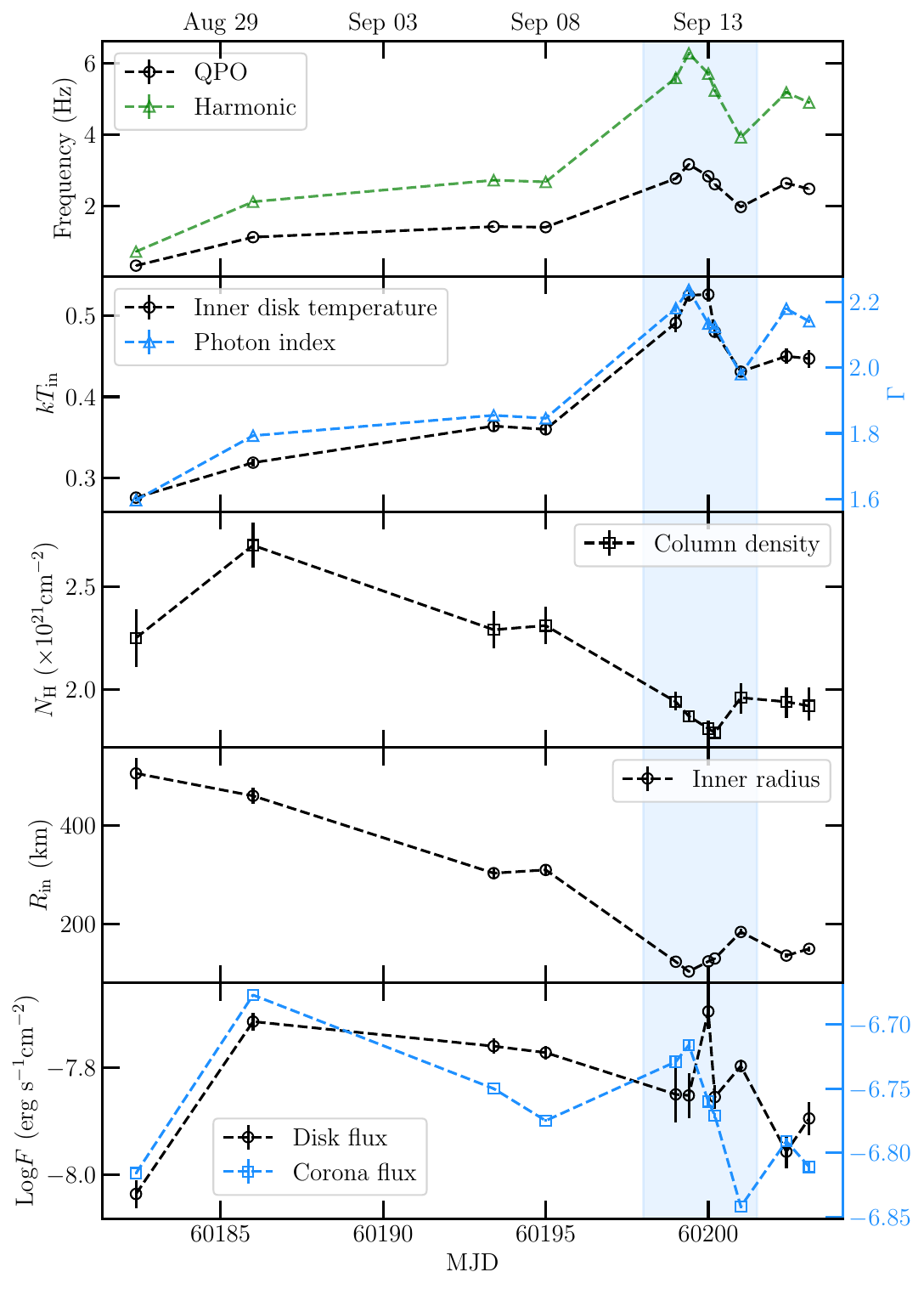}
    \caption{{Evolution of the QPO frequency and energy spectral properties. The flux is estimated using the \texttt{diskbb} and \texttt{nthcomp} models in the 1–10 keV energy range. The blue-shaded region indicates the period of the first flare.}}
    \label{fig:evo_curve}
\end{figure}

\subsection{The rms and phase-lag spectrum of type-C QPOs}\label{rms_lag}
In this study, we averaged the PDSs obtained from each orbit of every observation listed in Table \ref{tab:observations1} to ensure consistency and reliability in our subsequent discussions and qualitative analysis. The type-C QPO exhibits a centroid frequency that increases from 0.33 Hz to 2.63 Hz as it transitions from the normal state to the first flare. 
{In Fig.~\ref{fig:qpo},  we present six representative PDSs in the 1–10 keV energy range and show the fundamental QPO in the energy bands 1.0–1.5, 1.5–2.8, 2.8–5.0, 5.0–6.5, and 6.5–10 keV. These results reveal that the fundamental QPO is independent of energy in each observation interval during the first flare.}

Additionally, our findings in Table \ref{tab:observations1} show that the fractional rms increases during the rise phase, after which it remains relatively constant. As shown in Fig.~\ref{fig:rms_lag}, the evolution of rms in the normal state before MJD 60198 is consistent with the findings of \citet{Yu2024MNRAS}. The larger QPO rms amplitude observed above 6.5 keV suggests that the variability is driven by the corona rather than the disk. 
{Furthermore, we have plotted the phase lag within the 5–10 keV energy band as a function of time, using the 1–1.5 keV band as the reference, under the assumption that these type-C QPOs are driven by the same mechanism.}
Our findings show a general decreasing trend throughout the observation period, with a significant drop from -0.16 to -0.35, followed by a slight increase near MJD 60201.

{To investigate the energy dependence of the soft lag, we plot the phase lags for each energy sub-band in the right panel of Fig.~\ref{fig:rms_lag}. The resulting lag-energy spectra exhibit a characteristic semi-'U' shape during the flare.} The horizontal error bars represent the width of the energy channels. 
{\citet{2019ApJPatra} reported a correlation between the radio intensity and the amount of soft lag for a sample of outbursting Galactic black hole candidates. This correlation suggests that the observed soft lags might arise from the change of disk morphology due to the interaction with a strong outflow or jet. Based on this, the observed variation of the soft lag
(Figure \ref{fig:rms_lag}) can be interpreted as evidence of jet activity during its flare period in Swift J1727.8-1613.
Since soft phase lags are from disk reprocessing of hard photons from the corona \citep{Uttley2014A&ARv, Ingram2019MNRAS}, we surmise that the Comptonized region in Swift J1727.8-1613 is coupled to its jet activity in the flare period.}

\subsection{Co-evolution between spectral and timing properties}

The QPO frequency is found to be tightly correlated with the evolution of the energy spectrum in BHXBs. %During the outburst phase, some sources have shown tight correlations between the timing and spectral properties 
(e.g., \citealt{Vignarca2003A&A, Mendez2022NatAs,Zhang2022MNRAS,Fu2022RAA,Rawat2023MNRAS.524,Rawat2023MNRAS,Chatterjee2024arXiv}).
{In Fig. 4, we show the evolution of the QPO and energy spectral properties as Swift J1727.8–161 transitions into the flare state. The fundamental QPO frequency increases from 0.33 Hz to 2.63 Hz, with a bump peaking at 3.16 Hz during the flare, while the second harmonic emerges at a frequency roughly twice that of the fundamental QPO. In terms of spectral properties, the $\Gamma$ and $T_{\text{in}}$ keep the same increasing trend, but with a significant bump during the first flare period. However, the normalization of the \texttt{diskbb} model\footnote{\url{https://heasarc.gsfc.nasa.gov/docs/xanadu/xspec/manual/node165.html}} follows the opposite trend.} 

We derived the inner disk radius $R_{\rm in}$ from
${\rm norm} = (r_{\rm in}/D_{\rm 10})^{2}{\rm cos}\theta$,
where the norm is extracted from spectral fitting, $D_{10}$ represents the distance to the source in units of 10 kpc, $\theta$ denotes the angle of the disk, and $r_{\rm in}$ is the apparent inner disk radius which can be corrected to the true radius as $R_{\rm in}\approx \kappa^2\times\xi\times r_{\rm in}$. For Swift J1727.8–161, the values of $D_{10}=0.27$ (\citealt{Mata2024A&A}), $\theta=80^{\circ}$ (\citealt{Chatterjee2024arXiv}), $\kappa=1.7$ (\citealt{Shimura1995ApJ}) and $\xi=0.412$ (\citealt{Kubota1998PASJ}) have been utilized in the calculation. We have also estimated the flux from \texttt{diskbb} and \texttt{nthcomp}, as shown in Fig. \ref{fig:evo_curve}. 
{Notably, the coronal flux exhibits a significant decline around MJD 60201, suggesting a weakening of the corona.}

% We caution the $R_{\rm in}$ is the apparent inner disk radius instead of the true radius. We did not correct the apparent inner disk radius to the true radius for simplicity in this work (for information on the correction factor see e.g. \citealt{Kubota1998PASJ}). The $R_{\rm in}$ is derived from the normalization parameter of \texttt{diskbb} model, following the relation ${\rm norm} = (R_{\rm in}/D_{\rm 10})^{2}$cos$\theta$, where the norm is extracted from fitting the observed data, $D_{10}$ represents the distance to the source in units of 10 kpc, and $\theta$ denotes the angle of the disk. For Swift J1727.8–161, the values of $D_{10}=0.27$ (\citealt{Mata2024A&A}) and $\theta=80^{\circ}$ (\citealt{Chatterjee2024arXiv}) have been utilized in the calculation. 

\begin{figure}[ht!]
  \centering
  \begin{minipage}{0.49\linewidth}
    \centering
    \includegraphics[width=\linewidth]{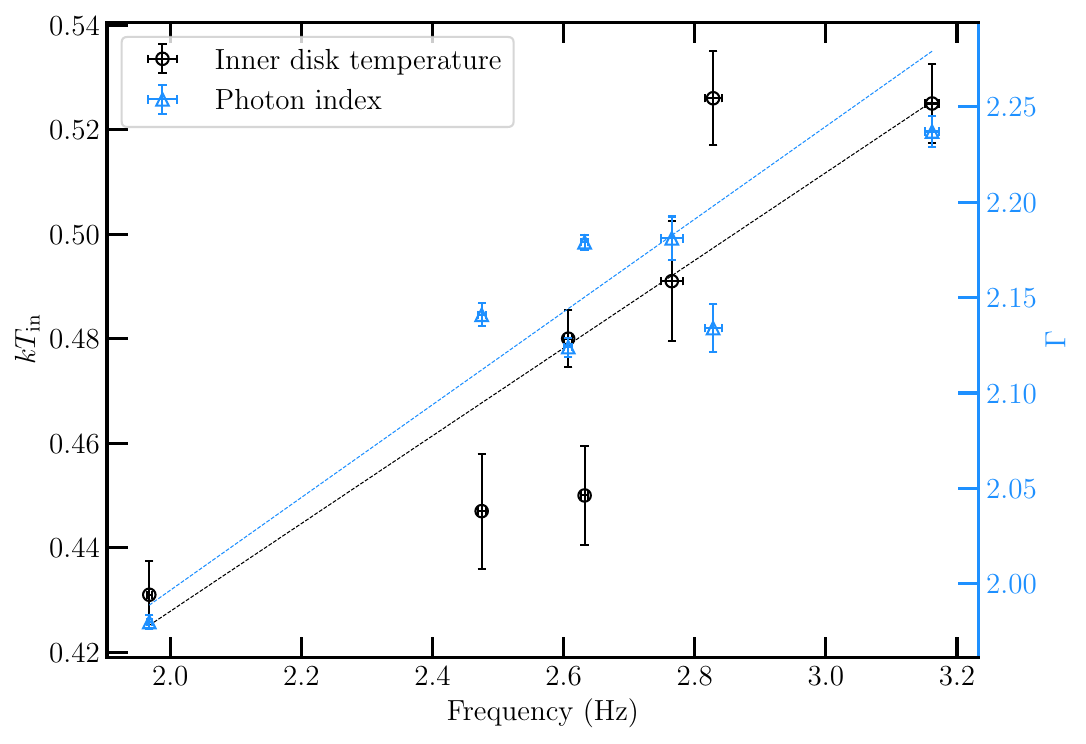}
  \end{minipage}
  \hfill
  \begin{minipage}{0.49\linewidth}
    \centering
    \includegraphics[width=\linewidth]{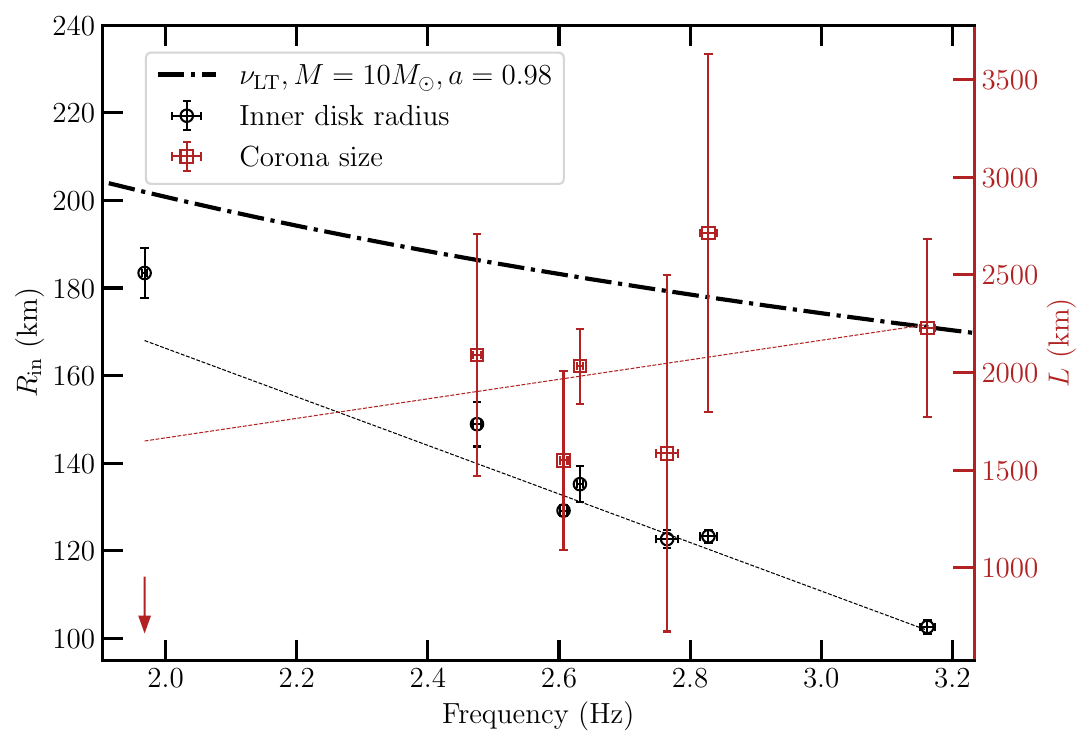}
  \end{minipage}
  
  \caption{
  % The relation between the energy spectral properties and the QPO frequency. The left panel shows the variations of the inner disk temperature $kT_{\rm in}$ (black circle) and photon index $\Gamma$ (blue triangles) with respect to the QPO frequency. The right panel shows the variations of the inner disk radius $R_{\rm in}$ (black circle) and corona size $L$ (red square) with respect to the QPO frequency.
  The variation with QPO frequency of (left) the inner disk temperature $kT_{\rm in}$ (black circles) and photon index $\Gamma$ (blue triangles) and (right) the inner disk radius $R_{\rm in}$ (black circles) and corona size $L$ (red squares).}
  \label{fig:qpo_corr}
\end{figure}

It is evident that there is a strong correlation between QPO frequency and spectral physical parameters during the flare. To further investigate possible co-evolution between the QPO and disk/corona, we present the variations of the disk and corona properties (i.e., inner disk temperature $kT_{\rm in}$, photon index of power-law $\Gamma$, inner disk radius $R_{\rm in}$ and corona size $L$) as functions of the QPO frequency in Fig. \ref{fig:qpo_corr}. As the QPO frequency increases from 1.97 Hz to $\sim$ 3 Hz, the inner disk temperature increases from 0.43 keV to $\sim$ 0.48 keV. The photon index and corona size exhibit a similar trend as the inner disk temperature. In contrast, the inner disk radius shows an opposite trend compared to the other parameters, i.e., the inner edge of the disk moves towards the black hole as the QPO frequency increases. We further calculate the Pearson correlation coefficients and the p-values to validate these correlations. All correlations were statistically significant (p-value$< 0.05$), except for the corona size (p-value=0.70). Strong correlations ($|r|>0.8$) were also observed for most parameters, while the corona size showed a weak correlation ($r=0.2$). These results suggest a significant connection between QPOs and the accretion disk. As for corona, the Pearson correlation analysis fails to establish a significant correlation between QPO frequency and corona size. However, we can not rule out the possibility that this correlation is true given the observed relationship between QPO frequency and the photon index. The high p-value and low correlation coefficient can be attributed to three factors: the exclusion of the upper-limit data point from the correlation analysis, substantial measurement errors ($\sim$30 percent), and the limited sample size from a short time scale ($\sim$ 3 days). These factors significantly impact the statistical results.

The LFQPOs are explicable within the framework of the relativistic precession model (RPM; \citealt{Stella1998ApJ}), which has a geometrical origin. The QPO frequency in the RPM model is assumed to be the nodal precession or LT precession frequency, and the theoretical frequency is negatively correlated with the orbital radius as (e.g., \citealt{Ingram2009MNRAS})

\begin{equation}
    \nu_{\phi} = \frac{c}{2\pi R_{\rm g} (r^{3/2} + a)},
\end{equation}
\begin{equation}
    \nu_{\rm LT} = \nu_{\phi} \left( 1 - \sqrt{1 - \frac{4a}{r^{3/2}} + \frac{3a}{r^{2}}} \right),
\end{equation}
where $a$ is the dimensionless spin parameter, and $r$ is the orbital radius in units of gravitational radius $R_{\rm g}$. As shown in the right panel of Fig. \ref{fig:qpo_corr}, the inner disk radius decreases as the QPO frequency increases, following a similar trend to $\nu_{\rm LT}$ ($a=0.98$ from \citealt{Peng2024ApJL}). However, theoretical predictions suggest a larger radius at the same frequency compared to the observational measurements. \cite{Kubota2024MNRAS} found a tight anti-correlation between the corona outer radius (truncated radius) and the LFQPO frequency which is in remarkably good qualitative and quantitative agreement with the prediction of LT precession in \citet{Ingram2009MNRAS}. $R_{\rm in}$ represents the inner radius of the blackbody disk and is smaller than the corona outer radius. Therefore, the theoretical predictions of radius from the LT precession naturally exceed the observational results of $R_{\rm in}$ (black dash-dot line in the right panel of Fig. \ref{fig:qpo_corr}).  Besides, the broadly constant feedback fraction (left panel of Fig. \ref{fig:corona}, see the discussion in the next subsection) indicates that the outer radius of the corona varies quasi-synchronously with $R_{\rm in}$, suggesting an anti-correlation between the outer radius of the corona and QPO frequency. These results provide possible evidence that the LFQPOs observed in the first flare of Swift J1727.8-1613 are dominated by the Lense-Thirring precession.
% Additionally, the nearly constant feedback fraction (left panel of Fig. \ref{fig:corona}, see the discussion in the next subsection) indicates that the outer radius of the corona varies as $R$, suggesting an anti-correlation between the outer radius of the corona and QPO frequency.

We observe a cautiously positive correlation between the corona size and QPO frequency during a short period of the first flare of Swift J1727.8-1613, which could be partially consistent with the findings of \citet{Mendez2022NatAs,Garcia2022MNRAS}. They reported that the corona contracts to $\sim 100$ km as the QPO frequency decreases from $\sim6$ Hz to $\sim2$ Hz in GRS 1915 + 105, based on more than a decade of data. Furthermore, the right panel of Fig. \ref{fig:qpo_corr} shows that the corona size reaches its minimum at $\sim2$ Hz, while the QPO lag in the high-energy band is near zero (obs9 of Fig. \ref{fig:rms_lag}) during the flare of Swift J1727.8-1613. Considering the anti-correlation between the inner disk radius and QPO frequency, it has been suggested that the LT precession of the inner edge of the accretion disk could modulate the LFQPO frequency. If LT precession dominates the QPO in the flare state of Swift J1727.8-1613, a near-zero QPO lag at $\sim$ 2 Hz would imply that the corona size is approximately equal to the inner disk radius \citep{Mendez2022NatAs, Garcia2022MNRAS}. This implication aligns more or less with the result shown in the right panel of Fig. \ref{fig:qpo_corr}, where $L<950$ km and $R_{\rm in}=182$ km at a QPO frequency of approximately 2 Hz.

\subsection{Tracking the geometry of the Comptonized region}\label{sec:corona_jetlike}
% Based on the fitting results of the rms and lag spectra of the QPO using the \texttt{vkompthdk} model, along with the time-averaged spectrum, we propose a scenario of a temporarily extended jet-like corona during the first flare period in the HIMS of Swift J1727.8–1613. This analysis also incorporates the quasi-simultaneous radio observations reported by \citet{Peters2023ATel}.

%

\begin{figure}[ht!]
  \centering
  \begin{minipage}{0.49\linewidth}
    \centering
    \includegraphics[width=\linewidth]{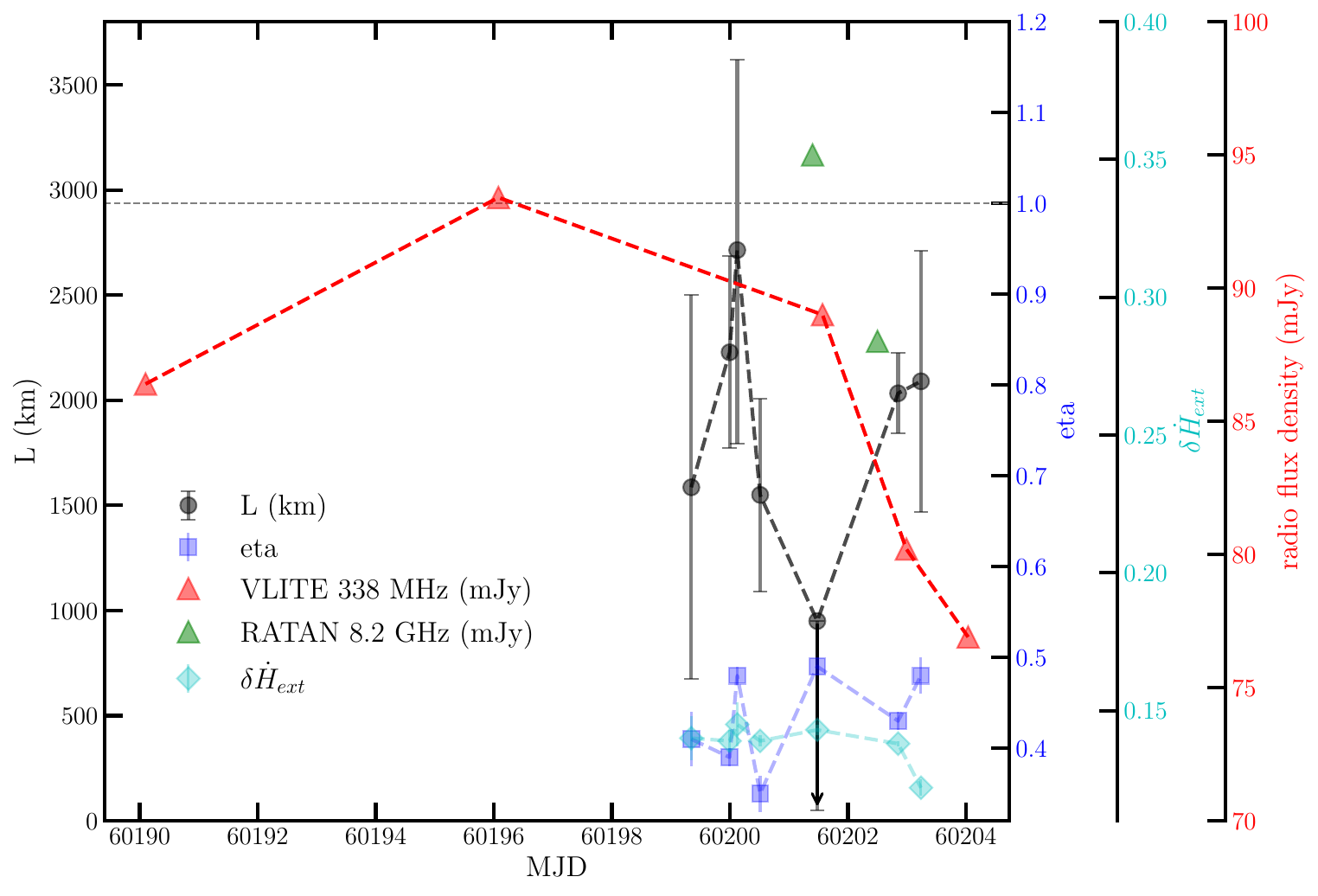}
  \end{minipage}
  \hfill
  \begin{minipage}{0.49\linewidth}
    \centering
    \includegraphics[width=\linewidth]{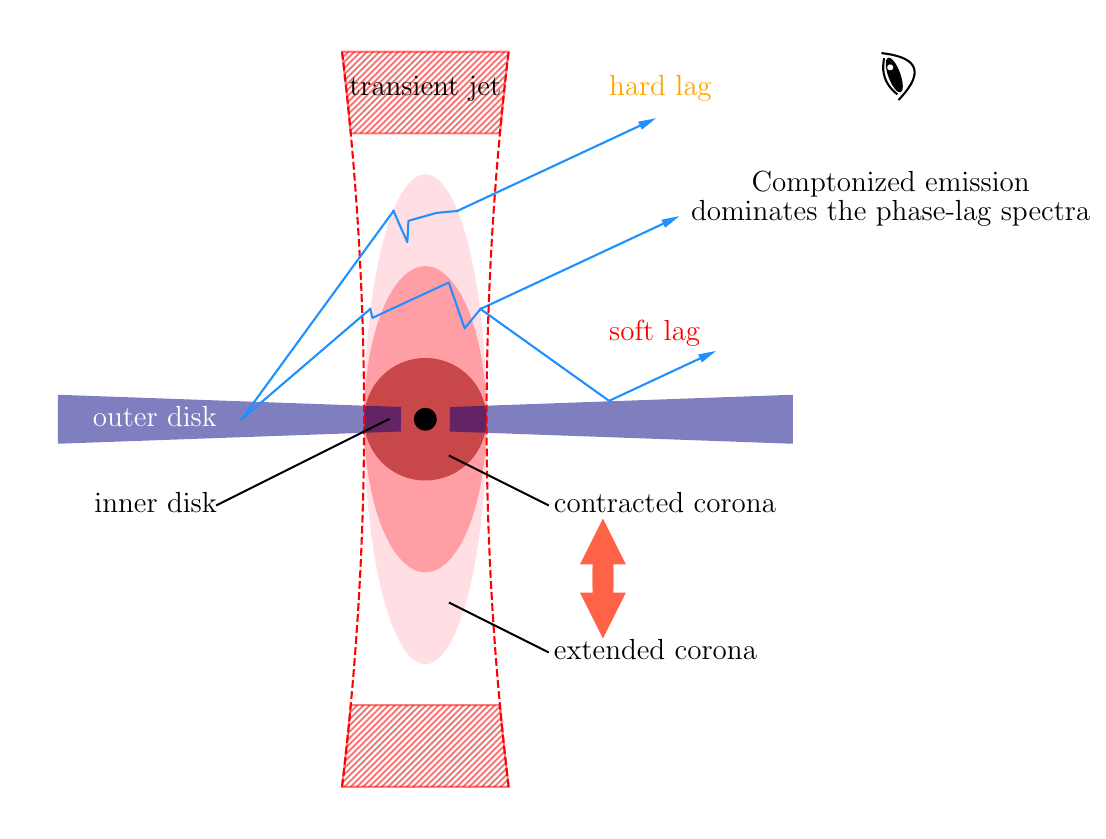}
  \end{minipage}

  \caption{The left panel shows the evolution of the corona size $L$ (black), the feedback fraction $\eta$ (blue), the VLITE radio flux (red), and the amplitude of the variability of the external heating rate (cyan) of Swift J1727.8–1613. The blue horizontal dashed line is the upper limit of 1, according to the definition of the time-dependent Comptonization model.
 The right panel represents the schematic picture of the corona evolution during the flare period in Swift J1727.8–1613. During the flare period, the corona expanded first and then contracted, with the transient jet quenched. The observed phase-lag spectra is dominated by the Comptonized emission, with soft lag driven by the reflection process of the accretion disk and hard lag by Compton scattering in the corona.}
  \label{fig:corona}
\end{figure}

{In Fig. \ref{fig:corona}, we illustrate the evolution of the corona size ($L$), the feedback fraction ($\eta$), and the amplitude of the variability of the external heating rate ($\delta\dot{H}_{ext}$), derived from the time-dependent model \texttt{vkompthdk} that fits the rms and phase-lag spectra.
During the first flare period, the size of the corona expands from $\sim$ 1580 km to $\sim$ 2710 km as the QPO approaches its peak, and then rapidly contracts to below 950 km as the QPO frequency drops to 1.97 Hz. 
Throughout this period, the feedback fraction which represents the fraction of photons scattered in the corona that return to the soft photon source \citep{Bellavita2022MNRAS}, remains broadly constant ranging from 0.3 to 0.5. This suggests that the corona maintains covering the accretion disk to some extent and the outer radius of the corona varies quasi-synchronously with $R_{\rm in}$ during the entire flare period, as shown in the right panel of Fig.~\ref{fig:corona}. 
Meanwhile, the amplitude of the variability of the external heating rate ($\delta\dot{H}_{ext}$) shows a constant value near 0.14. These findings suggest that the corona is not spherical but contracts perpendicularly to the accretion disk rather than horizontally. Besides, the LT radii ($\sim$ 100 km) in this work are well below the size of the corona (at the scale of $\sim$ 1000 km), providing additional support for the jet-like morphology of the Comptonized region.}

% In this work, the LT radii ($\sim$ 100 km) are well below the size of the corona (at the scale of $\sim$ 1000 km), indicating that the corona covers a relatively large fraction of the inner disk in Swift J1727.8-1613, which is consistent with the results of $\eta$ near 0.4.

% The LT radius \citep{Ingram2009MNRAS}, calculated from the QPO frequency assuming a 10 M$_{\odot}$ black hole with a spin of 0.998, decreases from 200 km to 100 km as the QPO frequency evolves from 2 Hz to 9 Hz \citep{Sridhar2019MNRAS}. 

Our results suggest a vertically extended corona, which aligns with several recent studies of different BHXBs.
\citet{harikrishna2022type} analyzed spectral-timing data of type-C to type-B QPOs in H 1743–322 and modelled a jet or a vertically extended, optically thick Comptonization region. Similarly, \citet{Liu2022ApJ} proposed a vertically extended corona at the jet base in MAXI J1348-630 to explain the disappearance and reappearance of the type-B QPO, invoking the Bardeen-Petterson effect. In the study of MAXI J1535-571, \citet{Zhang2023MNRAS} modelled the corona through type-B QPOs and found that the size of the jet-like corona extended vertically from approximately 3000 km to 6500 km in the intermediate state, which is comparable to the corona structure we observed in Swift J1727.8-1613. Furthermore, the jet-like corona can be supported by another approach. Using the reflection model \texttt{relxillCP} \citep{Garcia2014ApJ}, \citet{You2021Nature} found that the corona in MAXI J1820+070 outflows more rapidly as it moves closer to the black hole, suggesting a jet-like corona that gains energy as it outflows,

% Our results align with those of \citet{harikrishna2022type}, who analyzed spectral-timing data of type-C to type-B QPOs in H 1743–322 and modelled a jet or a vertically extended, optically thick Comptonization region. Similarly, \citet{Liu2022ApJ} proposed a vertically extended corona at the base of the jet in MAXI J1348-630 to explain the disappearance and reappearance of the type-B QPO, invoking the Bardeen-Petterson effect. In the study of MAXI J1535-571, \citet{Zhang2023MNRAS} modelled the corona through type-B QPOs and found that the size of the jet-like corona extended vertically from approximately 3000 km to 6500 km in the intermediate state. This finding is comparable to the corona structure we observed in Swift J1727.8-1613 and can be further supported by various approaches. For instance, using the reflection model \texttt{relxillCP} \citep{Garcia2014ApJ} to fit Insight-HXMT data, \citet{You2021Nature} found that the corona in MAXI J1820+070 outflows more rapidly as it moves closer to the black hole, suggesting a jet-like corona that gains energy as it outflows.

Next, we compare our results with the radio emissions observed from VLITE by \citet{Peters2023ATel} for this source during the period of the HIMS to SIMS transition. If the first and second flares are driven by the same mechanism and the radio luminosity $L_{\rm R}$ and X-ray luminosity $L_{\rm X}$ generally follow the non-linear relation $L_{\rm R} \propto {L_{\rm X}}^\beta$, it is plausible that the peak of the first flare should be around MJD 60199, aligning with the MAXI observations. To support this, we have added two sample points at 8.2 GHz from RATAN in Fig. \ref{fig:corona}, which strengthens our speculation.
This is reasonable given the coherence between the radio flux density and the X-ray light curve observed by RATAN and MAXI, respectively, during the subsequent flare and quenching (details can be seen in \citealt{Ingram2023arXiv,Miller-Jones2023ATel,Peters2023ATel}). Essentially, the radio emissions during the transition period from MJD 60190 to MJD 60204 are more likely indicative of a transient jet being launched, which is bright and consists of discrete relativistic ejecta from the black hole \citep{Miller-Jones2012MNRAS, Russell2019ApJ}. {Based on this, the significant decline in coronal flux around MJD 60201 can be interpreted as a result of the disruption or fragmentation of the corona when the transient jet launched.}
It is noted that the radio emission is suppressed rapidly near MJD 60202, a similar behavior has been observed by \citet{Mendez2022NatAs}, i.e., a low radio emission at or above a QPO frequency of $\sim$ 2.0 Hz. In this case, our results showing a temporarily shrinking corona after a radio brightening or an ejection are similar to observations of a contracting corona due to a quenching jet during the outburst of MAXI J1820+070 \citep{kara2019corona}. Subsequently, the corona as the jet base has recovered to its original size, which is also reasonable due to material spiraling into the corona along the magnetic field as the inner disk radius decreases \citep{Galeev1979ApJ, Haardt1991ApJ}. We note that the observed connection between the corona and the jet aligns with the findings of \citet{Wang2021ApJ} and \citet{Zhang2023MNRAS}, which reported transient jet emergence following the expansion of the corona. However, this result remains tentative due to the low sampling rate of VLITE during the first flare.

Based on the results and discussion presented in this section, we propose the following evolutionary scenario for the disk-corona-jet system during the first flare period (depicted in the right panel of Fig. \ref{fig:corona}). Initially, the corona expands vertically to its maximum size as the QPO frequency approaches its peak value, while the inner radius of the disk reaches a local minimum. The corona then contracts rapidly in the perpendicular direction, likely due to episodic jet launching (i.e., part of the corona is converted into a jet), as detected by VLITE and RATAN. Subsequently, the corona recovers as both the QPO frequency and the inner disk radius increase and leads up to the onset of the next flare.
In this scenario, the inner disk is much smaller than the corona's size ($\sim$1000 km), hence we didn't depict its variation in the right panel of Fig. \ref{fig:corona}. We predict that the corona continues to envelop the inner regions of the accretion disk, exhibiting a jet-like morphology throughout the flare state until the corona softens. The outer radius of the corona and the inner region of the disk move toward or away from the black hole in a synchronized manner. Consequently, photons from the variable jet-like corona in the vertical direction irradiate the disk, which reprocesses and re-emits the radiation, which results in corresponding changes in negative QPO lags (see Section \ref{rms_lag}).

% {\bf QPO at 2 Hz are consistent with the scenario described in \citealt{Mendez2022NatAs}, in which the energy that is initially stored in the X-ray corona is gradually released into the radio jet and, quite possibly, the X-ray corona itself becomes the jet}

\subsection{Phenomenological remarks}
We have speculated on the presence of a truncated radius using two different methods. As shown in the right panel of Fig.~\ref{fig:qpo_corr}, the inner radius and QPO curves predicted from the Lense-Thirring geometrical effect are higher than those derived from the actual observations. Additionally, the feedback fraction remains broadly constant, ranging from 0.30 to 0.50, suggesting that the corona likely maintains partial coverage of the accretion disk throughout the entire flare period, which resembles an advection-dominated accretion flow (ADAF). Thus, we consider the corona to be somewhat akin to the ADAF in the onward discussions. Recent research by \citet{Jiang2024MNRAStmp} proposed a physical model for the radio and X-ray correlation in BHXBs, which may also aid in understanding the evolving corona structure of Swift J1727.8-1613 during the first flare observed in this study.

For simplicity, we present a toy model similar to the conventional ADAF case. The value of the critical accretion rate $\dot{M}_{\rm cr}$ for the accretion mode transition can be estimated by equating the ion-electron equilibration timescale with the accretion timescale in the normal ADAF case \citep{Narayan1995ApJ, Narayan1998tbha}, which leads to
\begin{equation}
    \dot{M}_{\mathrm{cr}}=(1+f_{\mathrm{m}})^2\dot{M}_{\mathrm{cr},0}=\alpha^2(1+f_{\mathrm{m}})^2\dot{M}_{\mathrm{Edd}},
\end{equation}
where 
% \begin{equation}
%     f_\mathrm{m}(1+f_\mathrm{m})^{-1}=9.15\times10^{-19}m\dot{m}_\mathrm{cr}^{-1}r^{5/2}B_z^2,
% \end{equation}
\begin{equation}
    \dot{M}_{\mathrm{Edd}}\equiv \frac{L_{\mathrm{Edd}}}{0.1c^2}=\frac{4\pi GM_{\rm BH}}{0.1c\kappa_T},
\end{equation}
$M_{\rm BH}$ is the black hole mass, 
% $m$ is $M_{\rm BH}/M_{\odot}$, $\dot{m}_{cr}$ is $\dot{M}_{\rm cr}/\dot{M}_{\rm Edd}$, $B_z$ is the vertical components of the large-scale magnetic field at the disc surface,
and $\dot{M}_{\mathrm{cr},0}$ is the commonly used critical accretion rate. For an ADAF with magnetic outflows, its luminosity can be approximated as $L_{\rm ADAF}^{'} \sim (1+f_m)L_{\rm ADAF}$, which means the ADAF with magnetically driven outflows is $\sim f_{\rm m}$ times more luminous than the conventional ADAF accreting at the critical rate. According to \citet{Jiang2024MNRAStmp}, the generated field strength $B_{\rm pd}(r_{\rm tr})$ of the large-scale field of the outer disc at the truncated radius $r_{\rm tr}$ can be calculated by 
\begin{equation}
    B_{\rm pd}(r_{\rm tr}) \sim 2.48 \times 10^8 \alpha^{-1/20} m^{-11/20} \dot{m}_{\rm d}^{3/5} r_{\rm tr}^{-49/40} {\rm Gauss}.
\end{equation}

The elevated soft X-ray flux observed during the first flare, resulting from small perturbations in the accretion rate, is assumed to primarily originate from the ADAF with magnetic outflows. A constant viscosity parameter $\alpha$ was utilized over this short timescale, with $\dot{m}=\dot{m}_{\rm d} \sim \dot{m}_{\rm cr}$. Additionally, the accretion rate $\dot{m}$ was estimated using $\dot{m}_{\rm cr}(r) \sim {T_e}^{3/2}(r)\alpha^2$. In this equation, the electron temperature $T_e$ varies with radius as a power law in the outer region of the ADAF, while the temperature distribution remains relatively flat in the inner region near the black hole \citep{Manmoto2000ApJ, Xu2010ApJ}. 

% \textbf{Therefore, the morphology of the corona may expand either radially or vertically, accompanied by variations in the magnetic field along the corresponding direction, as the critical accretion rate changes.}

Based on the relationship $B_{\rm pd} \sim \dot{m}_{\rm d}^{3/5}$, it can be inferred that the magnetic field perpendicular to the disk temporarily strengthens with the increased accretion rate during brief episodes as in the first flare period. Consequently, the radio jet produced by synchrotron radiation is expected to diminish as the magnetic field weakens with the decline in the accretion rate.
This speculation is partially supported by \citet{Bouchet2024arXiv}, who observed a peak in the polarization fraction near MJD 60201 from the soft $\gamma$--ray data of INTEGRAL/IBIS for Swift J1727.8-1613 during its outburst. The polarization results in the soft $\gamma$--ray band (250-300 keV) are presented in Table \ref{tab:polarization}. This polarization data provides evidence that the magnetic field plays a pivotal role in modulating the coupling of the disk-corona-jet, as discussed in Section \ref{sec:corona_jetlike}. 
Further investigation using additional facilities is necessary to explore the connection between accretion rate and magnetic polarization. Instruments such as Insight-HXMT and NuSTAR, which cover a wider energy range, and IXPE, which provides polarization data, will help validate this toy model for Swift J1727.8-1613, particularly during its transition into the flare state.

% Thus, the inner ADAF with magnetic outflows might expand radially or vertically, with a corresponding variable magnetic field in the radial or vertical direction.

Even though we have performed a simple phenomenological analysis in this work, partly supported by compiled data from polarization observations, the detailed physics within a real corona of BHXBs remains unclear. More observational data from upcoming X-ray missions with larger area detectors and polarization capabilities (e.g., eXTP; \citealt{Zhang2019SCPMA}), as well as GRMHD simulations \citep{Narayan2012MNRAS, 2025ApJSridhar}, are required to provide a comprehensive understanding of the complete picture of the corona-jet evolution and the role of the magnetic field during the outburst of BHXBs.

% \begin{table}
    
% 	\centering
%         \caption{Polarization results in 250-300 KeV from \citet{Bouchet2024arXiv}.}
% 	\label{tab:polarization}
% 	\begin{tabular}{cccccc} % four columns, alignment for each
%             \hline
%             \hline
% 		Data group & 1 & 2 & 3 & 4 &5\\
% 		\hline
% 		Start (MJD) & 60181.8 & 60186.6 & 60194.2 & 60200.2 & 60207.4 \\
% 		Stop (MJD) & 60182.4 & 60193.0 & 60198.3 & 60203.7 & 60215.7\\
%             PA & - & $-51.2 \pm 3.2$ & $-47.4 \pm 6.0$ & $-43.4 \pm 5.8 $ & - \\
%             PF & $ <25 $ & $46 \pm 5$ & $50 \pm 10$ & $81 \pm 17$ & $<43$ \\
%             S/N & 29.7 & 55.3 & 43.5 & 26.1 & 20.3 \\
%             \hline
% 	\end{tabular}
% \end{table}

\begin{deluxetable*}{cccccc}
\tablenum{4}
\tablecaption{Polarization results in soft $\gamma$ ray band 250-300 KeV of INTEGRAL/IBIS from \citet{Bouchet2024arXiv}. \label{tab:polarization}}
\tablewidth{0pt}
\tablehead{
\colhead{Data group} & \colhead{1} & \colhead{2} & \colhead{3}& \colhead{4}& \colhead{5}
}
% \decimalcolnumbers
\startdata
Start (MJD) & 60181.8 & 60186.6 & 60194.2 & 60200.2 & 60207.4 \\
Stop (MJD) & 60182.4 & 60193.0 & 60198.3 & 60203.7 & 60215.7\\
PA & - & $-51.2 \pm 3.2^\circ$ & $-47.4 \pm 6.0^\circ$ & $-43.4 \pm 5.8^\circ $ & - \\
PF & $ <25 $ & $46 \pm 5$ & $50 \pm 10$ & $81 \pm 17$ & $<43$ \\
S/N & 29.7 & 55.3 & 43.5 & 26.1 & 20.3
\enddata
% \tablecomments{}
\end{deluxetable*}

\section{Summary}\label{sec:summary}

We analyzed the data from 11 NICER observations during the state transition of Swift J1727.8–1613 from the LHS to the flare state. We fit the energy spectra of the source, as well as the rms and lag spectra of the type-C QPO in this source, using the one-component time-dependent Comptonization model \texttt{vkompthdk}. Below, we summarize our results:

(1) A type-C QPO with a centroid frequency increasing from 0.33 Hz to 2.63 Hz was detected. During the first flare, the central frequency of the QPOs showed a significant elevation, accompanied by a sudden increase in the soft X-ray flux. This correlation supports a close connection between the QPO frequency variations and the inner accretion disk.

(2) Strong correlations were found between the QPO and the inner disk properties, providing possible evidence that the type-C QPOs observed in the flare period are modulated by Lense-Thirring precession.

(3) The data can be effectively fitted using the time-dependent Comptonization model \texttt{vkompthdk}. Based on this model, we have proposed a possible scenario for the evolution of the vertical jet-like corona geometry during the first flare period (a short time scale, $\sim$ 3 days), as derived from NICER observations and previously published radio data (see Section \ref{sec:corona_jetlike} for details). This evolution provides insights into the dynamic changes in the corona structure and its relationship with the observed QPO characteristics and radio emissions.

(4) We have also conducted a phenomenological analysis of the corona scenario in Swift J1727.8-1613 during the flare period, which is partially supported by polarization data from INTEGRAL/IBIS. However, the intrinsic physical mechanism under the variable corona remains unclear and requires further research.

\section*{Acknowledgements}

We thank the anonymous referee for insightful comments and useful suggestions that improved the paper. This work was supported by the National Key R\&D Intergovernmental Cooperation Program of China (grant No. 2023YFE0102300), the National SKA Program of China (grant Nos. 2022SKA0120102 and 2020SKA0120300), the National Key R\&D Program of China (grant No. 2021YFA0718500), the CAS `Light of West China' Program (grant No. 2021-XBQNXZ-005), and the NSFC (grant Nos. U2031212, 12233002, and 12025301). LC acknowledges the support from the Tianshan Talent Training Program (grant No. 2023TSYCCX0099). NC, YFH, TA, LCH, and AT acknowledge the support from the Xinjiang Tianchi Talent Program. LCH was supported by the National Key R\&D Program of China (2022YFF0503401), the NSFC (grant Nos. 11991052, 12233001), and the China Manned Space Project (CMS-CSST-2021-A04, CMS-CSST-2021-A06). This work was also partly supported by the Urumqi Nanshan Astronomy and Deep Space Exploration Observation and Research Station of Xinjiang (XJYWZ2303). 

%%
%% pdflatex sample631.tex
%% bibtext sample631
%% pdflatex sample631.tex
%% pdflatex sample631.tex

\bibliography{sample631}{}

\begin{thebibliography}{}
\expandafter\ifx\csname natexlab\endcsname\relax\def\natexlab#1{#1}\fi
\providecommand{\url}[1]{\href{#1}{#1}}
\providecommand{\dodoi}[1]{doi:~\href{http://doi.org/#1}{\nolinkurl{#1}}}
\providecommand{\doeprint}[1]{\href{http://ascl.net/#1}{\nolinkurl{http://ascl.net/#1}}}
\providecommand{\doarXiv}[1]{\href{https://arxiv.org/abs/#1}{\nolinkurl{https://arxiv.org/abs/#1}}}

\bibitem[{{Alabarta} {et~al.}(2021){Alabarta}, {Altamirano}, {M{\'e}ndez}, {C{\'u}neo}, {Vincentelli}, {Castro-Segura}, {Garc{\'\i}a}, {Luff}, \& {Veledina}}]{Alabarta2021MNRAS}
{Alabarta}, K., {Altamirano}, D., {M{\'e}ndez}, M., {et~al.} 2021, \mnras, 507, 5507, \dodoi{10.1093/mnras/stab2241}

\bibitem[{{Ar{\'e}valo} \& {Uttley}(2006)}]{Arevalo2006MNRAS}
{Ar{\'e}valo}, P., \& {Uttley}, P. 2006, \mnras, 367, 801, \dodoi{10.1111/j.1365-2966.2006.09989.x}

\bibitem[{{Bachetti} {et~al.}(2015){Bachetti}, {Harrison}, {Cook}, {Tomsick}, {Schmid}, {Grefenstette}, {Barret}, {Boggs}, {Christensen}, {Craig}, {Fabian}, {F{\"u}rst}, {Gandhi}, {Hailey}, {Kara}, {Maccarone}, {Miller}, {Pottschmidt}, {Stern}, {Uttley}, {Walton}, {Wilms}, \& {Zhang}}]{Bachetti2015ApJ}
{Bachetti}, M., {Harrison}, F.~A., {Cook}, R., {et~al.} 2015, \apj, 800, 109, \dodoi{10.1088/0004-637X/800/2/109}

\bibitem[{{Bachetti} {et~al.}(2023){Bachetti}, {Huppenkothen}, {Khan}, {Mishra}, {Stevens}, {Sharma}, {Swinbank}, {Desai}, {Rashid}, {Martinez Ribeiro}, {Tripathi}, {Sip{\H{o}}cz}, {Vats}, {Tappina}, {Mastroserio}, {Omargamal8}, {Davis}, {Rasquinha}, {Balm}, {Mumford}, {Shukla}, {Campana}, {Parkma99}, {Garg}, {Tandon}, {Hota}, {Anand}, {Nick}, {Raj}, \& {Mishra}}]{Bachetti2023stingray}
{Bachetti}, M., {Huppenkothen}, D., {Khan}, U., {et~al.} 2023, {StingraySoftware/stingray: Version 1.1.2}, v1.1.2,  Zenodo, \dodoi{10.5281/zenodo.7970570}

\bibitem[{{Bambi} {et~al.}(2021){Bambi}, {Brenneman}, {Dauser}, {Garc{\'\i}a}, {Grinberg}, {Ingram}, {Jiang}, {Liu}, {Lohfink}, {Marinucci}, {Mastroserio}, {Middei}, {Nampalliwar}, {Nied{\'z}wiecki}, {Steiner}, {Tripathi}, \& {Zdziarski}}]{Bambi2021SSRv}
{Bambi}, C., {Brenneman}, L.~W., {Dauser}, T., {et~al.} 2021, \ssr, 217, 65, \dodoi{10.1007/s11214-021-00841-8}

\bibitem[{{Basko} {et~al.}(1974){Basko}, {Sunyaev}, \& {Titarchuk}}]{Basko1974A&A}
{Basko}, M.~M., {Sunyaev}, R.~A., \& {Titarchuk}, L.~G. 1974, \aap, 31, 249

\bibitem[{{Bellavita} {et~al.}(2022){Bellavita}, {Garc{\'\i}a}, {M{\'e}ndez}, \& {Karpouzas}}]{Bellavita2022MNRAS}
{Bellavita}, C., {Garc{\'\i}a}, F., {M{\'e}ndez}, M., \& {Karpouzas}, K. 2022, \mnras, 515, 2099, \dodoi{10.1093/mnras/stac1922}

\bibitem[{{Belloni} \& {Hasinger}(1990)}]{Belloni1990A&A}
{Belloni}, T., \& {Hasinger}, G. 1990, \aap, 227, L33

\bibitem[{{Belloni} {et~al.}(2005){Belloni}, {Homan}, {Casella}, {van der Klis}, {Nespoli}, {Lewin}, {Miller}, \& {M{\'e}ndez}}]{Belloni2005A&A}
{Belloni}, T., {Homan}, J., {Casella}, P., {et~al.} 2005, \aap, 440, 207, \dodoi{10.1051/0004-6361:20042457}

\bibitem[{{Belloni} {et~al.}(2002){Belloni}, {Psaltis}, \& {van der Klis}}]{Belloni2002ApJ}
{Belloni}, T., {Psaltis}, D., \& {van der Klis}, M. 2002, \apj, 572, 392, \dodoi{10.1086/340290}

\bibitem[{{Bouchet} {et~al.}(2024){Bouchet}, {Rodriguez}, {Cangemi}, {Thalhammer}, {Laurent}, {Grinberg}, {Wilms}, \& {Pottschimdt}}]{Bouchet2024arXiv}
{Bouchet}, T., {Rodriguez}, J., {Cangemi}, F., {et~al.} 2024, arXiv e-prints, arXiv:2407.05871, \dodoi{10.48550/arXiv.2407.05871}

\bibitem[{{Bu} {et~al.}(2015){Bu}, {Chen}, {Li}, {Qu}, {Belloni}, \& {Zhang}}]{Bu2015ApJ}
{Bu}, Q.-c., {Chen}, L., {Li}, Z.-s., {et~al.} 2015, \apj, 799, 2, \dodoi{10.1088/0004-637X/799/1/2}

\bibitem[{{Casella} {et~al.}(2005){Casella}, {Belloni}, \& {Stella}}]{Casella2005ApJ}
{Casella}, P., {Belloni}, T., \& {Stella}, L. 2005, \apj, 629, 403, \dodoi{10.1086/431174}

\bibitem[{{Castro-Tirado} {et~al.}(2023){Castro-Tirado}, {Sanchez-Ramirez}, {Caballero-Garcia}, {Perez-Garcia}, {Fernandez-Garcia}, {Guziy}, {Hu}, {Blazek}, {Hermelo}, {Pinter}, {Meintjes}, {van Heerden}, {Martin-Carrillo}, {Hanlon}, {Hiriart}, {Lee}, {Carrasco-Garcia}, {Park}, {Gritsevich}, {Castellon}, {Perez del Pulgar}, \& {Reina}}]{Castro-Tirado2023ATel}
{Castro-Tirado}, A.~J., {Sanchez-Ramirez}, R., {Caballero-Garcia}, M.~D., {et~al.} 2023, The Astronomer's Telegram, 16208, 1

\bibitem[{{Chatterjee} {et~al.}(2024){Chatterjee}, {Mondal}, {Singh}, \& {Sugizaki}}]{Chatterjee2024arXiv}
{Chatterjee}, K., {Mondal}, S., {Singh}, C.~B., \& {Sugizaki}, M. 2024, arXiv e-prints, arXiv:2405.01498, \dodoi{10.48550/arXiv.2405.01498}

\bibitem[{{Corbel} {et~al.}(2004){Corbel}, {Fender}, {Tomsick}, {Tzioumis}, \& {Tingay}}]{Corbel2004ApJ}
{Corbel}, S., {Fender}, R.~P., {Tomsick}, J.~A., {Tzioumis}, A.~K., \& {Tingay}, S. 2004, \apj, 617, 1272, \dodoi{10.1086/425650}

\bibitem[{{Cui} {et~al.}(1997){Cui}, {Zhang}, {Focke}, \& {Swank}}]{Cui1997ApJ}
{Cui}, W., {Zhang}, S.~N., {Focke}, W., \& {Swank}, J.~H. 1997, \apj, 484, 383, \dodoi{10.1086/304341}

\bibitem[{{Done} {et~al.}(2007){Done}, {Gierli{\'n}ski}, \& {Kubota}}]{Done2007A&ARv}
{Done}, C., {Gierli{\'n}ski}, M., \& {Kubota}, A. 2007, \aapr, 15, 1, \dodoi{10.1007/s00159-007-0006-1}

\bibitem[{{Esin} {et~al.}(1997){Esin}, {McClintock}, \& {Narayan}}]{Esin1997ApJ}
{Esin}, A.~A., {McClintock}, J.~E., \& {Narayan}, R. 1997, \apj, 489, 865, \dodoi{10.1086/304829}

\bibitem[{{Fender}(2006)}]{Fender2006csxs.book}
{Fender}, R. 2006, in Compact stellar X-ray sources, ed. W.~H.~G. {Lewin} \& M.~{van der Klis}, Vol.~39, 381--419, \dodoi{10.48550/arXiv.astro-ph/0303339}

\bibitem[{{Fender}(2001)}]{Fender2001MNRAS}
{Fender}, R.~P. 2001, \mnras, 322, 31, \dodoi{10.1046/j.1365-8711.2001.04080.x}

\bibitem[{{Fender} {et~al.}(2004){Fender}, {Belloni}, \& {Gallo}}]{Fender2004MNRAS}
{Fender}, R.~P., {Belloni}, T.~M., \& {Gallo}, E. 2004, \mnras, 355, 1105, \dodoi{10.1111/j.1365-2966.2004.08384.x}

\bibitem[{{Fu} {et~al.}(2022){Fu}, {Song}, {Ding}, {Zhang}, {Qu}, {Zhang}, {Zhang}, {Bu}, {Huang}, {Ma}, {Yang}, {Tuo}, {Lu}, {Zhou}, {Wu}, {Li}, \& {Xu}}]{Fu2022RAA}
{Fu}, Y.-C., {Song}, L.~M., {Ding}, G.~Q., {et~al.} 2022, Research in Astronomy and Astrophysics, 22, 115002, \dodoi{10.1088/1674-4527/ac8d80}

\bibitem[{{Galeev} {et~al.}(1979){Galeev}, {Rosner}, \& {Vaiana}}]{Galeev1979ApJ}
{Galeev}, A.~A., {Rosner}, R., \& {Vaiana}, G.~S. 1979, \apj, 229, 318, \dodoi{10.1086/156957}

\bibitem[{{Gallo} {et~al.}(2003){Gallo}, {Fender}, \& {Pooley}}]{Gallo2003MNRAS}
{Gallo}, E., {Fender}, R.~P., \& {Pooley}, G.~G. 2003, \mnras, 344, 60, \dodoi{10.1046/j.1365-8711.2003.06791.x}

\bibitem[{{Garc{\'\i}a} {et~al.}(2022){Garc{\'\i}a}, {Karpouzas}, {M{\'e}ndez}, {Zhang}, {Zhang}, {Belloni}, \& {Altamirano}}]{Garcia2022MNRAS}
{Garc{\'\i}a}, F., {Karpouzas}, K., {M{\'e}ndez}, M., {et~al.} 2022, \mnras, 513, 4196, \dodoi{10.1093/mnras/stac1202}

\bibitem[{{Garc{\'\i}a} {et~al.}(2021){Garc{\'\i}a}, {M{\'e}ndez}, {Karpouzas}, {Belloni}, {Zhang}, \& {Altamirano}}]{Garcia2021MNRAS}
{Garc{\'\i}a}, F., {M{\'e}ndez}, M., {Karpouzas}, K., {et~al.} 2021, \mnras, 501, 3173, \dodoi{10.1093/mnras/staa3944}

\bibitem[{{Garc{\'\i}a} {et~al.}(2014){Garc{\'\i}a}, {Dauser}, {Lohfink}, {Kallman}, {Steiner}, {McClintock}, {Brenneman}, {Wilms}, {Eikmann}, {Reynolds}, \& {Tombesi}}]{Garcia2014ApJ}
{Garc{\'\i}a}, J., {Dauser}, T., {Lohfink}, A., {et~al.} 2014, \apj, 782, 76, \dodoi{10.1088/0004-637X/782/2/76}

\bibitem[{Gendreau {et~al.}(2016)Gendreau, Arzoumanian, Adkins, Albert, Anders, Aylward, Baker, Balsamo, Bamford, Benegalrao, {et~al.}}]{gendreau2016neutron}
Gendreau, K.~C., Arzoumanian, Z., Adkins, P.~W., {et~al.} 2016, in Space telescopes and instrumentation 2016: Ultraviolet to gamma ray, Vol. 9905, SPIE, 420--435

\bibitem[{{George} \& {Fabian}(1991)}]{George1991MNRAS}
{George}, I.~M., \& {Fabian}, A.~C. 1991, \mnras, 249, 352, \dodoi{10.1093/mnras/249.2.352}

\bibitem[{{Gilfanov}(2010)}]{Gilfanov2010}
{Gilfanov}, M. 2010, in Lecture Notes in Physics, Berlin Springer Verlag, ed. T.~{Belloni}, Vol. 794, 17, \dodoi{10.1007/978-3-540-76937-8_2}

\bibitem[{{Haardt} \& {Maraschi}(1991)}]{Haardt1991ApJ}
{Haardt}, F., \& {Maraschi}, L. 1991, \apjl, 380, L51, \dodoi{10.1086/186171}

\bibitem[{{Harikrishna} \& {Sriram}(2022)}]{harikrishna2022type}
{Harikrishna}, S., \& {Sriram}, K. 2022, \mnras, 516, 5148, \dodoi{10.1093/mnras/stac2527}

\bibitem[{{Homan} \& {Belloni}(2005)}]{Homan2005Ap&SS}
{Homan}, J., \& {Belloni}, T. 2005, \apss, 300, 107, \dodoi{10.1007/s10509-005-1197-4}

\bibitem[{{Homan} {et~al.}(2001){Homan}, {Wijnands}, {van der Klis}, {Belloni}, {van Paradijs}, {Klein-Wolt}, {Fender}, \& {M{\'e}ndez}}]{Homan2001ApJS}
{Homan}, J., {Wijnands}, R., {van der Klis}, M., {et~al.} 2001, \apjs, 132, 377, \dodoi{10.1086/318954}

\bibitem[{{Huppenkothen} {et~al.}(2019{\natexlab{a}}){Huppenkothen}, {Bachetti}, {Stevens}, {Migliari}, {Balm}, {Hammad}, {Khan}, {Mishra}, {Rashid}, {Sharma}, {Ribeiro}, \& {Blanco}}]{Huppenkothen2019stingraya}
{Huppenkothen}, D., {Bachetti}, M., {Stevens}, A., {et~al.} 2019{\natexlab{a}}, The Journal of Open Source Software, 4, 1393, \dodoi{10.21105/joss.01393}

\bibitem[{{Huppenkothen} {et~al.}(2019{\natexlab{b}}){Huppenkothen}, {Bachetti}, {Stevens}, {Migliari}, {Balm}, {Hammad}, {Khan}, {Mishra}, {Rashid}, {Sharma}, {Martinez Ribeiro}, \& {Valles Blanco}}]{Huppenkothen2019stingrayb}
{Huppenkothen}, D., {Bachetti}, M., {Stevens}, A.~L., {et~al.} 2019{\natexlab{b}}, \apj, 881, 39, \dodoi{10.3847/1538-4357/ab258d}

\bibitem[{{Ingram} \& {Done}(2012)}]{Ingram2012MNRAS}
{Ingram}, A., \& {Done}, C. 2012, \mnras, 419, 2369, \dodoi{10.1111/j.1365-2966.2011.19885.x}

\bibitem[{{Ingram} {et~al.}(2009){Ingram}, {Done}, \& {Fragile}}]{Ingram2009MNRAS}
{Ingram}, A., {Done}, C., \& {Fragile}, P.~C. 2009, \mnras, 397, L101, \dodoi{10.1111/j.1745-3933.2009.00693.x}

\bibitem[{{Ingram} {et~al.}(2019){Ingram}, {Mastroserio}, {Dauser}, {Hovenkamp}, {van der Klis}, \& {Garc{\'\i}a}}]{Ingram2019MNRAS}
{Ingram}, A., {Mastroserio}, G., {Dauser}, T., {et~al.} 2019, \mnras, 488, 324, \dodoi{10.1093/mnras/stz1720}

\bibitem[{{Ingram} \& {van der Klis}(2013)}]{Ingram2013MNRAS}
{Ingram}, A., \& {van der Klis}, M. 2013, \mnras, 434, 1476, \dodoi{10.1093/mnras/stt1107}

\bibitem[{{Ingram} {et~al.}(2023){Ingram}, {Bollemeijer}, {Veledina}, {Dovciak}, {Poutanen}, {Egron}, {Russell}, {Trushkin}, {Negro}, {Ratheesh}, {Capitanio}, {Connors}, {Neilsen}, {Kraus}, {Noemi Iacolina}, {Pellizzoni}, {Pilia}, {Carotenuto}, {Matt}, {Mastroserio}, {Kaaret}, {Bianchi}, {Garcia}, {Bachetti}, {Wu}, {Costa}, {Ewing}, {Kravtsov}, {Krawczynski}, {Loktev}, {Marinucci}, {Marra}, {Mikusincova}, {Nathan}, {Parra}, {Petrucci}, {Righini}, {Soffitta}, {Steiner}, {Svoboda}, {Tombesi}, {Tugliani}, {Ursini}, {Yang}, {Zane}, {Zhang}, {Agudo}, {Antonelli}, {Baldini}, {Baumgartner}, {Bellazzini}, {Bongiorno}, {Bonino}, {Brez}, {Bucciantini}, {Castellano}, {Cavazzuti}, {Chen}, {Ciprini}, {De Rosa}, {Del Monte}, {Di Gesu}, {Di Lalla}, {Di Marco}, {Donnarumma}, {Doroshenko}, {Ehlert}, {Enoto}, {Evangelista}, {Fabiani}, {Ferrazzoli}, {Gunji}, {Hayashida}, {Heyl}, {Iwakiri}, {Jorstad}, {Karas}, {Kislat}, {Kitaguchi}, {Kolodziejczak}, {La Monaca}, {Latronico}, {Liodakis}, {Maldera}, {Manfreda}, {Marin},
  {Marscher}, {Marshall}, {Massaro}, {Mitsuishi}, {Mizuno}, {Muleri}, {Ng}, {O'Dell}, {Omodei}, {Oppedisano}, {Papitto}, {Pavlov}, {Peirson}, {Perri}, {Pesce-Rollins}, {Possenti}, {Puccetti}, {Ramsey}, {Rankin}, {Roberts}, {Romani}, {Sgro}, {Slane}, {Spandre}, {Swartz}, {Tamagawa}, {Tavecchio}, {Taverna}, {Tawara}, {Tennant}, {Thomas}, {Trois}, {Tsygankov}, {Turolla}, {Vink}, {Weisskopf}, \& {Xie}}]{Ingram2023arXiv}
{Ingram}, A., {Bollemeijer}, N., {Veledina}, A., {et~al.} 2023, arXiv e-prints, arXiv:2311.05497, \dodoi{10.48550/arXiv.2311.05497}

\bibitem[{{Ingram} {et~al.}(2024){Ingram}, {Bollemeijer}, {Veledina}, {Dov{\v{c}}iak}, {Poutanen}, {Egron}, {Russell}, {Trushkin}, {Negro}, {Ratheesh}, {Capitanio}, {Connors}, {Neilsen}, {Kraus}, {Iacolina}, {Pellizzoni}, {Pilia}, {Carotenuto}, {Matt}, {Mastroserio}, {Kaaret}, {Bianchi}, {Garc{\'\i}a}, {Bachetti}, {Wu}, {Costa}, {Ewing}, {Kravtsov}, {Krawczynski}, {Loktev}, {Marinucci}, {Marra}, {Miku{\v{s}}incov{\'a}}, {Nathan}, {Parra}, {Petrucci}, {Righini}, {Soffitta}, {Steiner}, {Svoboda}, {Tombesi}, {Tugliani}, {Ursini}, {Yang}, {Zane}, {Zhang}, {Agudo}, {Antonelli}, {Baldini}, {Baumgartner}, {Bellazzini}, {Bongiorno}, {Bonino}, {Brez}, {Bucciantini}, {Castellano}, {Cavazzuti}, {Chen}, {Ciprini}, {De Rosa}, {Del Monte}, {Di Gesu}, {Di Lalla}, {Di Marco}, {Donnarumma}, {Doroshenko}, {Ehlert}, {Enoto}, {Evangelista}, {Fabiani}, {Ferrazzoli}, {Gunji}, {Hayashida}, {Heyl}, {Iwakiri}, {Jorstad}, {Karas}, {Kislat}, {Kitaguchi}, {Kolodziejczak}, {La Monaca}, {Latronico}, {Liodakis}, {Maldera}, {Manfreda},
  {Marin}, {Marscher}, {Marshall}, {Massaro}, {Mitsuishi}, {Mizuno}, {Muleri}, {Ng}, {O'Dell}, {Omodei}, {Oppedisano}, {Papitto}, {Pavlov}, {Peirson}, {Perri}, {Pesce-Rollins}, {Possenti}, {Puccetti}, {Ramsey}, {Rankin}, {Roberts}, {Romani}, {Sgr{\`o}}, {Slane}, {Spandre}, {Swartz}, {Tamagawa}, {Tavecchio}, {Taverna}, {Tawara}, {Tennant}, {Thomas}, {Trois}, {Tsygankov}, {Turolla}, {Vink}, {Weisskopf}, {Xie}, \& {IXPE Collaboration}}]{2024ApJIngram}
---. 2024, \apj, 968, 76, \dodoi{10.3847/1538-4357/ad3faf}

\bibitem[{{Ingram} \& {Motta}(2019)}]{Ingram2019NewAR}
{Ingram}, A.~R., \& {Motta}, S.~E. 2019, \nar, 85, 101524, \dodoi{10.1016/j.newar.2020.101524}

\bibitem[{{Jiang} {et~al.}(2024){Jiang}, {Li}, {Cao}, {You}, {Zdziarski}, \& {Xu}}]{Jiang2024MNRAStmp}
{Jiang}, Y., {Li}, S., {Cao}, X., {et~al.} 2024, \mnras, \dodoi{10.1093/mnras/stae1777}

\bibitem[{Kara {et~al.}(2019)Kara, Steiner, Fabian, Cackett, Uttley, Remillard, Gendreau, Arzoumanian, Altamirano, Eikenberry, {et~al.}}]{kara2019corona}
Kara, E., Steiner, J., Fabian, A., {et~al.} 2019, Nature, 565, 198

\bibitem[{{Karpouzas} {et~al.}(2020){Karpouzas}, {M{\'e}ndez}, {Ribeiro}, {Altamirano}, {Blaes}, \& {Garc{\'\i}a}}]{Karpouzas2020MNRAS}
{Karpouzas}, K., {M{\'e}ndez}, M., {Ribeiro}, E.~M., {et~al.} 2020, \mnras, 492, 1399, \dodoi{10.1093/mnras/stz3502}

\bibitem[{{Kubota} {et~al.}(2024){Kubota}, {Done}, {Tsurumi}, \& {Mizukawa}}]{Kubota2024MNRAS}
{Kubota}, A., {Done}, C., {Tsurumi}, K., \& {Mizukawa}, R. 2024, \mnras, 528, 1668, \dodoi{10.1093/mnras/stae067}

\bibitem[{{Kubota} {et~al.}(1998){Kubota}, {Tanaka}, {Makishima}, {Ueda}, {Dotani}, {Inoue}, \& {Yamaoka}}]{Kubota1998PASJ}
{Kubota}, A., {Tanaka}, Y., {Makishima}, K., {et~al.} 1998, \pasj, 50, 667, \dodoi{10.1093/pasj/50.6.667}

\bibitem[{{Liu} {et~al.}(2022){Liu}, {Huang}, {Bu}, {Yu}, {Yang}, {Zhang}, {Kong}, {Xiao}, {Qu}, {Zhang}, {Zhang}, {Song}, {Jia}, {Ma}, {Tao}, {Ge}, {Liu}, {Yan}, {Ma}, {Ren}, {Zhou}, {Li}, {Wu}, {Xu}, {Du}, {Fu}, {Xiao}, {Ding}, \& {Yu}}]{Liu2022ApJ}
{Liu}, H.~X., {Huang}, Y., {Bu}, Q.~C., {et~al.} 2022, \apj, 938, 108, \dodoi{10.3847/1538-4357/ac88c6}

\bibitem[{{Ma} {et~al.}(2023){Ma}, {M{\'e}ndez}, {Garc{\'\i}a}, {Sai}, {Zhang}, \& {Zhang}}]{ma2023variable}
{Ma}, R., {M{\'e}ndez}, M., {Garc{\'\i}a}, F., {et~al.} 2023, \mnras, 525, 854, \dodoi{10.1093/mnras/stad2284}

\bibitem[{{Ma} {et~al.}(2021){Ma}, {Tao}, {Zhang}, {Zhang}, {Bu}, {Ge}, {Chen}, {Qu}, {Zhang}, {Lu}, {Song}, {Yang}, {Yuan}, {Cai}, {Cao}, {Chang}, {Chen}, {Chen}, {Chen}, {Chen}, {Chen}, {Cui}, {Cui}, {Deng}, {Dong}, {Du}, {Fu}, {Gao}, {Gao}, {Gao}, {Gu}, {Guan}, {Guo}, {Han}, {Huang}, {Huo}, {Ji}, {Jia}, {Jiang}, {Jiang}, {Jin}, {Jin}, {Kong}, {Li}, {Li}, {Li}, {Li}, {Li}, {Li}, {Li}, {Li}, {Li}, {Li}, {Li}, {Liang}, {Liao}, {Liu}, {Liu}, {Liu}, {Liu}, {Liu}, {Liu}, {Lu}, {Lu}, {Luo}, {Luo}, {Meng}, {Nang}, {Nie}, {Ou}, {Sai}, {Shang}, {Song}, {Sun}, {Tan}, {Tuo}, {Wang}, {Wang}, {Wang}, {Wang}, {Wang}, {Wang}, {Wen}, {Wu}, {Wu}, {Wu}, {Xiao}, {Xiao}, {Xie}, {Xiong}, {Xu}, {Xu}, {Yang}, {Yang}, {Yang}, {Yi}, {Yin}, {You}, {Zhang}, {Zhang}, {Zhang}, {Zhang}, {Zhang}, {Zhang}, {Zhang}, {Zhang}, {Zhang}, {Zhang}, {Zhang}, {Zhang}, {Zhang}, {Zhang}, {Zhang}, {Zhang}, {Zhao}, {Zhao}, {Zheng}, {Zhou}, {Zhou}, {Zhu}, {Zhu}, \& {Zhuang}}]{Ma2021NatAs}
{Ma}, X., {Tao}, L., {Zhang}, S.-N., {et~al.} 2021, Nature Astronomy, 5, 94, \dodoi{10.1038/s41550-020-1192-2}

\bibitem[{{Manmoto}(2000)}]{Manmoto2000ApJ}
{Manmoto}, T. 2000, \apj, 534, 734, \dodoi{10.1086/308768}

\bibitem[{{Markoff} {et~al.}(2005){Markoff}, {Nowak}, \& {Wilms}}]{Markoff2005ApJ}
{Markoff}, S., {Nowak}, M.~A., \& {Wilms}, J. 2005, \apj, 635, 1203, \dodoi{10.1086/497628}

\bibitem[{{Mata S{\'a}nchez} {et~al.}(2024){Mata S{\'a}nchez}, {Mu{\~n}oz-Darias}, {Armas Padilla}, {Casares}, \& {Torres}}]{Mata2024A&A}
{Mata S{\'a}nchez}, D., {Mu{\~n}oz-Darias}, T., {Armas Padilla}, M., {Casares}, J., \& {Torres}, M.~A.~P. 2024, \aap, 682, L1, \dodoi{10.1051/0004-6361/202348754}

\bibitem[{{M{\'e}ndez} {et~al.}(2022){M{\'e}ndez}, {Karpouzas}, {Garc{\'\i}a}, {Zhang}, {Zhang}, {Belloni}, \& {Altamirano}}]{Mendez2022NatAs}
{M{\'e}ndez}, M., {Karpouzas}, K., {Garc{\'\i}a}, F., {et~al.} 2022, Nature Astronomy, 6, 577, \dodoi{10.1038/s41550-022-01617-y}

\bibitem[{{M{\'e}ndez} \& {van der Klis}(1997)}]{Mendez1997ApJ}
{M{\'e}ndez}, M., \& {van der Klis}, M. 1997, \apj, 479, 926, \dodoi{10.1086/303914}

\bibitem[{{Miller-Jones} {et~al.}(2023){Miller-Jones}, {Bahramian}, {Altamirano}, {Homan}, {Russell}, \& {Sivakoff}}]{Miller-Jones2023ATel}
{Miller-Jones}, J.~C.~A., {Bahramian}, A., {Altamirano}, D., {et~al.} 2023, The Astronomer's Telegram, 16271, 1

\bibitem[{{Miller-Jones} {et~al.}(2012){Miller-Jones}, {Sivakoff}, {Altamirano}, {Coriat}, {Corbel}, {Dhawan}, {Krimm}, {Remillard}, {Rupen}, {Russell}, {Fender}, {Heinz}, {K{\"o}rding}, {Maitra}, {Markoff}, {Migliari}, {Sarazin}, \& {Tudose}}]{Miller-Jones2012MNRAS}
{Miller-Jones}, J.~C.~A., {Sivakoff}, G.~R., {Altamirano}, D., {et~al.} 2012, \mnras, 421, 468, \dodoi{10.1111/j.1365-2966.2011.20326.x}

\bibitem[{{Mirabel} \& {Rodr{\'\i}guez}(1994)}]{Mirabel1994Nature}
{Mirabel}, I.~F., \& {Rodr{\'\i}guez}, L.~F. 1994, \nat, 371, 46, \dodoi{10.1038/371046a0}

\bibitem[{Mitsuda {et~al.}(1984)Mitsuda, Inoue, Koyama, Makishima, Matsuoka, Ogawara, Shibazaki, Suzuki, Tanaka, \& Hirano}]{mitsuda1984energy}
Mitsuda, K., Inoue, H., Koyama, K., {et~al.} 1984, Astronomical Society of Japan, Publications (ISSN 0004-6264), vol. 36, no. 4, 1984, p. 741-759., 36, 741

\bibitem[{{Miyamoto} {et~al.}(1991){Miyamoto}, {Kimura}, {Kitamoto}, {Dotani}, \& {Ebisawa}}]{Miyamoto1991ApJ}
{Miyamoto}, S., {Kimura}, K., {Kitamoto}, S., {Dotani}, T., \& {Ebisawa}, K. 1991, \apj, 383, 784, \dodoi{10.1086/170837}

\bibitem[{{Miyamoto} \& {Kitamoto}(1989)}]{Miyamoto1989Nature}
{Miyamoto}, S., \& {Kitamoto}, S. 1989, \nat, 342, 773, \dodoi{10.1038/342773a0}

\bibitem[{{Motta} {et~al.}(2012){Motta}, {Homan}, {Mu{\~n}oz Darias}, {Casella}, {Belloni}, {Hiemstra}, \& {M{\'e}ndez}}]{Motta2012MNRAS}
{Motta}, S., {Homan}, J., {Mu{\~n}oz Darias}, T., {et~al.} 2012, \mnras, 427, 595, \dodoi{10.1111/j.1365-2966.2012.22037.x}

\bibitem[{{Nakajima} {et~al.}(2023){Nakajima}, {Negoro}, {Serino}, {Mihara}, {Kobayashi}, {Tanaka}, {Soejima}, {Kudo}, {Kawamuro}, {Yamada}, {Tamagawa}, {Kawai}, {Matsuoka}, {Sakamoto}, {Sugita}, {Hiramatsu}, {Nishikawa}, {Yoshida}, {Tsuboi}, {Urabe}, {Nawa}, {Nemoto}, {Shidatsu}, {Takahashi}, {Niwano}, {Sato}, {Higuchi}, {Yatsu}, {Nakahira}, {Ueno}, {Tomida}, {Ishikawa}, {Ogawa}, {Kurihara}, {Ueda}, {Setoguchi}, {Yoshitake}, {Nakatani}, {Yamauchi}, {Hagiwara}, {Umeki}, {Otsuki}, {Yamaoka}, {Kawakubo}, {Sugizaki}, \& {Iwakiri}}]{Nakajima2023ATel}
{Nakajima}, M., {Negoro}, H., {Serino}, M., {et~al.} 2023, The Astronomer's Telegram, 16206, 1

\bibitem[{{Narayan} {et~al.}(1998){Narayan}, {Mahadevan}, \& {Quataert}}]{Narayan1998tbha}
{Narayan}, R., {Mahadevan}, R., \& {Quataert}, E. 1998, in Theory of Black Hole Accretion Disks, ed. M.~A. {Abramowicz}, G.~{Bj{\"o}rnsson}, \& J.~E. {Pringle}, 148--182, \dodoi{10.48550/arXiv.astro-ph/9803141}

\bibitem[{{Narayan} {et~al.}(2012){Narayan}, {S{\"A} dowski}, {Penna}, \& {Kulkarni}}]{Narayan2012MNRAS}
{Narayan}, R., {S{\"A} dowski}, A., {Penna}, R.~F., \& {Kulkarni}, A.~K. 2012, \mnras, 426, 3241, \dodoi{10.1111/j.1365-2966.2012.22002.x}

\bibitem[{{Narayan} \& {Yi}(1995)}]{Narayan1995ApJ}
{Narayan}, R., \& {Yi}, I. 1995, \apj, 452, 710, \dodoi{10.1086/176343}

\bibitem[{{Negoro} {et~al.}(2023){Negoro}, {Serino}, {Nakajima}, {Kobayashi}, {Tanaka}, {Soejima}, {Kudo}, {Mihara}, {Kawamuro}, {Yamada}, {Tamagawa}, {Kawai}, {Matsuoka}, {Sakamoto}, {Sugita}, {Hiramatsu}, {Nishikawa}, {Yoshida}, {Tsuboi}, {Urabe}, {Nawa}, {Nemoto}, {Shidatsu}, {Takahashi}, {Niwano}, {Sato}, {Higuchi}, {Yatsu}, {Nakahira}, {Ueno}, {Tomida}, {Ishikawa}, {Ogawa}, {Kurihara}, {Ueda}, {Setoguchi}, {Yoshitake}, {Nakatani}, {Yamauchi}, {Hagiwara}, {Umeki}, {Otsuki}, {Yamaoka}, {Kawakubo}, {Sugizaki}, \& {Iwakiri}}]{2023ATel16205....1N}
{Negoro}, H., {Serino}, M., {Nakajima}, M., {et~al.} 2023, The Astronomer's Telegram, 16205, 1

\bibitem[{{Nowak} {et~al.}(1999){Nowak}, {Vaughan}, {Wilms}, {Dove}, \& {Begelman}}]{Nowak1999ApJ}
{Nowak}, M.~A., {Vaughan}, B.~A., {Wilms}, J., {Dove}, J.~B., \& {Begelman}, M.~C. 1999, \apj, 510, 874, \dodoi{10.1086/306610}

\bibitem[{{O'Connor} {et~al.}(2023){O'Connor}, {Hare}, {Younes}, {Gendreau}, {Arzoumanian}, \& {Ferrara}}]{OConnor2023ATel}
{O'Connor}, B., {Hare}, J., {Younes}, G., {et~al.} 2023, The Astronomer's Telegram, 16207, 1

\bibitem[{{Page} {et~al.}(2023){Page}, {Dichiara}, {Gropp}, {Krimm}, {Parsotan}, {Williams}, \& {Neil Gehrels Swift Observatory Team}}]{Page2023GCN}
{Page}, K.~L., {Dichiara}, S., {Gropp}, J.~D., {et~al.} 2023, GRB Coordinates Network, 34537, 1

\bibitem[{{Patra} {et~al.}(2019){Patra}, {Chatterjee}, {Dutta}, {Chakrabarti}, \& {Nandi}}]{2019ApJPatra}
{Patra}, D., {Chatterjee}, A., {Dutta}, B.~G., {Chakrabarti}, S.~K., \& {Nandi}, P. 2019, \apj, 886, 137, \dodoi{10.3847/1538-4357/ab4c34}

\bibitem[{{Peirano} {et~al.}(2023){Peirano}, {M{\'e}ndez}, {Garc{\'\i}a}, \& {Belloni}}]{Peirano2023MNRAS}
{Peirano}, V., {M{\'e}ndez}, M., {Garc{\'\i}a}, F., \& {Belloni}, T. 2023, \mnras, 519, 1336, \dodoi{10.1093/mnras/stac3553}

\bibitem[{{Peng} {et~al.}(2024){Peng}, {Zhang}, {Shui}, {Zhang}, {Kong}, {Chen}, {Wang}, {Ji}, {Qu}, {Tao}, {Ge}, {Chang}, {Li}, {Li}, {Yu}, \& {Yan}}]{Peng2024ApJL}
{Peng}, J.-Q., {Zhang}, S., {Shui}, Q.-C., {et~al.} 2024, \apjl, 960, L17, \dodoi{10.3847/2041-8213/ad17ca}

\bibitem[{{Peters} {et~al.}(2023){Peters}, {Polisensky}, {Clarke}, {Giacintucci}, \& {Kassim}}]{Peters2023ATel}
{Peters}, W.~M., {Polisensky}, E., {Clarke}, T.~E., {Giacintucci}, S., \& {Kassim}, N.~E. 2023, The Astronomer's Telegram, 16279, 1

\bibitem[{{Rawat} {et~al.}(2023{\natexlab{a}}){Rawat}, {Husain}, \& {Misra}}]{Rawat2023MNRAS.524}
{Rawat}, D., {Husain}, N., \& {Misra}, R. 2023{\natexlab{a}}, \mnras, 524, 5869, \dodoi{10.1093/mnras/stad2220}

\bibitem[{{Rawat} {et~al.}(2023{\natexlab{b}}){Rawat}, {M{\'e}ndez}, {Garc{\'\i}a}, {Altamirano}, {Karpouzas}, {Zhang}, {Alabarta}, {Belloni}, {Jain}, \& {Bellavita}}]{Rawat2023MNRAS}
{Rawat}, D., {M{\'e}ndez}, M., {Garc{\'\i}a}, F., {et~al.} 2023{\natexlab{b}}, \mnras, 520, 113, \dodoi{10.1093/mnras/stad126}

\bibitem[{{Remillard} \& {McClintock}(2006)}]{Remillard2006ARA&A}
{Remillard}, R.~A., \& {McClintock}, J.~E. 2006, \araa, 44, 49, \dodoi{10.1146/annurev.astro.44.051905.092532}

\bibitem[{{Remillard} {et~al.}(2002){Remillard}, {Sobczak}, {Muno}, \& {McClintock}}]{Remillard2002ApJ}
{Remillard}, R.~A., {Sobczak}, G.~J., {Muno}, M.~P., \& {McClintock}, J.~E. 2002, \apj, 564, 962, \dodoi{10.1086/324276}

\bibitem[{{Russell} {et~al.}(2011){Russell}, {Miller-Jones}, {Maccarone}, {Yang}, {Fender}, \& {Lewis}}]{Russell2011ApJ}
{Russell}, D.~M., {Miller-Jones}, J.~C.~A., {Maccarone}, T.~J., {et~al.} 2011, \apjl, 739, L19, \dodoi{10.1088/2041-8205/739/1/L19}

\bibitem[{{Russell} {et~al.}(2019){Russell}, {Tetarenko}, {Miller-Jones}, {Sivakoff}, {Parikh}, {Rapisarda}, {Wijnands}, {Corbel}, {Tremou}, {Altamirano}, {Baglio}, {Ceccobello}, {Degenaar}, {van den Eijnden}, {Fender}, {Heywood}, {Krimm}, {Lucchini}, {Markoff}, {Russell}, {Soria}, \& {Woudt}}]{Russell2019ApJ}
{Russell}, T.~D., {Tetarenko}, A.~J., {Miller-Jones}, J.~C.~A., {et~al.} 2019, \apj, 883, 198, \dodoi{10.3847/1538-4357/ab3d36}

\bibitem[{{Shakura} \& {Sunyaev}(1973)}]{Shakura1973A&A}
{Shakura}, N.~I., \& {Sunyaev}, R.~A. 1973, \aap, 24, 337

\bibitem[{{Shimura} \& {Takahara}(1995)}]{Shimura1995ApJ}
{Shimura}, T., \& {Takahara}, F. 1995, \apj, 445, 780, \dodoi{10.1086/175740}

\bibitem[{{Sridhar} {et~al.}(2025){Sridhar}, {Ripperda}, {Sironi}, {Davelaar}, \& {Beloborodov}}]{2025ApJSridhar}
{Sridhar}, N., {Ripperda}, B., {Sironi}, L., {Davelaar}, J., \& {Beloborodov}, A.~M. 2025, \apj, 979, 199, \dodoi{10.3847/1538-4357/ada385}

\bibitem[{{Stella} \& {Vietri}(1998)}]{Stella1998ApJ}
{Stella}, L., \& {Vietri}, M. 1998, \apjl, 492, L59, \dodoi{10.1086/311075}

\bibitem[{{Uttley} {et~al.}(2014{\natexlab{a}}){Uttley}, {Cackett}, {Fabian}, {Kara}, \& {Wilkins}}]{Uttley2014A&ARv}
{Uttley}, P., {Cackett}, E.~M., {Fabian}, A.~C., {Kara}, E., \& {Wilkins}, D.~R. 2014{\natexlab{a}}, \aapr, 22, 72, \dodoi{10.1007/s00159-014-0072-0}

\bibitem[{{Uttley} {et~al.}(2014{\natexlab{b}}){Uttley}, {Cackett}, {Fabian}, {Kara}, \& {Wilkins}}]{2014A&ARvUttley}
---. 2014{\natexlab{b}}, \aapr, 22, 72, \dodoi{10.1007/s00159-014-0072-0}

\bibitem[{{van der Klis} {et~al.}(1987){van der Klis}, {Hasinger}, {Stella}, {Langmeier}, {van Paradijs}, \& {Lewin}}]{1987ApJvan}
{van der Klis}, M., {Hasinger}, G., {Stella}, L., {et~al.} 1987, \apjl, 319, L13, \dodoi{10.1086/184946}

\bibitem[{{Veledina} {et~al.}(2023){Veledina}, {Muleri}, {Dov{\v{c}}iak}, {Poutanen}, {Ratheesh}, {Capitanio}, {Matt}, {Soffitta}, {Tennant}, {Negro}, {Kaaret}, {Costa}, {Ingram}, {Svoboda}, {Krawczynski}, {Bianchi}, {Steiner}, {Garc{\'\i}a}, {Kravtsov}, {Nitindala}, {Ewing}, {Mastroserio}, {Marinucci}, {Ursini}, {Tombesi}, {Tsygankov}, {Yang}, {Weisskopf}, {Trushkin}, {Egron}, {Iacolina}, {Pilia}, {Marra}, {Miku{\v{s}}incov{\'a}}, {Nathan}, {Parra}, {Petrucci}, {Podgorn{\'y}}, {Tugliani}, {Zane}, {Zhang}, {Agudo}, {Antonelli}, {Bachetti}, {Baldini}, {Baumgartner}, {Bellazzini}, {Bongiorno}, {Bonino}, {Brez}, {Bucciantini}, {Castellano}, {Cavazzuti}, {Chen}, {Ciprini}, {De Rosa}, {Del Monte}, {Di Gesu}, {Di Lalla}, {Di Marco}, {Donnarumma}, {Doroshenko}, {Ehlert}, {Enoto}, {Evangelista}, {Fabiani}, {Ferrazzoli}, {Gunji}, {Hayashida}, {Heyl}, {Iwakiri}, {Jorstad}, {Karas}, {Kislat}, {Kitaguchi}, {Kolodziejczak}, {La Monaca}, {Latronico}, {Liodakis}, {Maldera}, {Manfreda}, {Marin}, {Marscher}, {Marshall},
  {Massaro}, {Mitsuishi}, {Mizuno}, {Ng}, {O'Dell}, {Omodei}, {Oppedisano}, {Papitto}, {Pavlov}, {Peirson}, {Perri}, {Pesce-Rollins}, {Possenti}, {Puccetti}, {Ramsey}, {Rankin}, {Roberts}, {Romani}, {Sgr{\`o}}, {Slane}, {Spandre}, {Swartz}, {Tamagawa}, {Tavecchio}, {Taverna}, {Tawara}, {Thomas}, {Trois}, {Turolla}, {Vink}, {Wu}, \& {Xie}}]{Veledina2023ApJ}
{Veledina}, A., {Muleri}, F., {Dov{\v{c}}iak}, M., {et~al.} 2023, \apjl, 958, L16, \dodoi{10.3847/2041-8213/ad0781}

\bibitem[{{Vignarca} {et~al.}(2003){Vignarca}, {Migliari}, {Belloni}, {Psaltis}, \& {van der Klis}}]{Vignarca2003A&A}
{Vignarca}, F., {Migliari}, S., {Belloni}, T., {Psaltis}, D., \& {van der Klis}, M. 2003, \aap, 397, 729, \dodoi{10.1051/0004-6361:20021542}

\bibitem[{{Wang} {et~al.}(2021){Wang}, {Mastroserio}, {Kara}, {Garc{\'\i}a}, {Ingram}, {Connors}, {van der Klis}, {Dauser}, {Steiner}, {Buisson}, {Homan}, {Lucchini}, {Fabian}, {Bright}, {Fender}, {Cackett}, \& {Remillard}}]{Wang2021ApJ}
{Wang}, J., {Mastroserio}, G., {Kara}, E., {et~al.} 2021, \apjl, 910, L3, \dodoi{10.3847/2041-8213/abec79}

\bibitem[{{Wijnands} {et~al.}(1999){Wijnands}, {Homan}, \& {van der Klis}}]{Wijnands1999ApJ}
{Wijnands}, R., {Homan}, J., \& {van der Klis}, M. 1999, \apjl, 526, L33, \dodoi{10.1086/312365}

\bibitem[{{Wood} {et~al.}(2024){Wood}, {Miller-Jones}, {Bahramian}, {Tingay}, {Prabu}, {Russell}, {Atri}, {Carotenuto}, {Altamirano}, {Motta}, {Hyland}, {Reynolds}, {Weston}, {Fender}, {K{\"o}rding}, {Maitra}, {Markoff}, {Migliari}, {Russell}, {Sarazin}, {Sivakoff}, {Soria}, {Tetarenko}, \& {Tudose}}]{2024ApJWood}
{Wood}, C.~M., {Miller-Jones}, J. C.~A., {Bahramian}, A., {et~al.} 2024, \apjl, 971, L9, \dodoi{10.3847/2041-8213/ad6572}

\bibitem[{{Xu} \& {Cao}(2010)}]{Xu2010ApJ}
{Xu}, Y.-D., \& {Cao}, X. 2010, \apj, 716, 1423, \dodoi{10.1088/0004-637X/716/2/1423}

\bibitem[{{Yang} {et~al.}(2023){Yang}, {Zhang}, {Zhang}, {M{\'e}ndez}, {Garc{\'\i}a}, {Huang}, {Bu}, {Liu}, {Yu}, {Wang}, {Tao}, {Altamirano}, {Qu}, {Zhang}, {Ma}, {Song}, {Jia}, {Ge}, {Liu}, {Yan}, {Li}, {Ren}, {Ma}, {Zhang}, {Xu}, {Ma}, {Du}, {Fu}, {Xiao}, {Li}, {Jin}, {Zhao}, \& {Zhao}}]{Yang2023MNRAS}
{Yang}, Z.-X., {Zhang}, L., {Zhang}, S.~N., {et~al.} 2023, \mnras, 521, 3570, \dodoi{10.1093/mnras/stad795}

\bibitem[{You {et~al.}(2021)}]{You2021Nature}
You, B., {et~al.} 2021, Nature Commun., 12, 1025, \dodoi{10.1038/s41467-021-21169-5}

\bibitem[{{Yu} {et~al.}(2024){Yu}, {Bu}, {Zhang}, {Liu}, {Zhang}, {Ducci}, {Tao}, {Santangelo}, {Doroshenko}, {Huang}, {Yang}, \& {Qu}}]{Yu2024MNRAS}
{Yu}, W., {Bu}, Q.-C., {Zhang}, S.-N., {et~al.} 2024, \mnras, 529, 4624, \dodoi{10.1093/mnras/stae835}

\bibitem[{{Zdziarski} {et~al.}(1996){Zdziarski}, {Johnson}, \& {Magdziarz}}]{Zdziarski1996MNRAS}
{Zdziarski}, A.~A., {Johnson}, W.~N., \& {Magdziarz}, P. 1996, \mnras, 283, 193, \dodoi{10.1093/mnras/283.1.193}

\bibitem[{{Zhang} {et~al.}(2019){Zhang}, {Santangelo}, {Feroci}, {Xu}, {Lu}, {Chen}, {Feng}, {Zhang}, {Brandt}, {Hernanz}, {Baldini}, {Bozzo}, {Campana}, {De Rosa}, {Dong}, {Evangelista}, {Karas}, {Meidinger}, {Meuris}, {Nandra}, {Pan}, {Pareschi}, {Orleanski}, {Huang}, {Schanne}, {Sironi}, {Spiga}, {Svoboda}, {Tagliaferri}, {Tenzer}, {Vacchi}, {Zane}, {Walton}, {Wang}, {Winter}, {Wu}, {in't Zand}, {Ahangarianabhari}, {Ambrosi}, {Ambrosino}, {Barbera}, {Basso}, {Bayer}, {Bellazzini}, {Bellutti}, {Bertucci}, {Bertuccio}, {Borghi}, {Cao}, {Cadoux}, {Campana}, {Ceraudo}, {Chen}, {Chen}, {Chevenez}, {Civitani}, {Cui}, {Cui}, {Dauser}, {Del Monte}, {Di Cosimo}, {Diebold}, {Doroshenko}, {Dovciak}, {Du}, {Ducci}, {Fan}, {Favre}, {Fuschino}, {G{\'a}lvez}, {Gao}, {Ge}, {Gevin}, {Grassi}, {Gu}, {Gu}, {Han}, {Hong}, {Hu}, {Ji}, {Jia}, {Jiang}, {Kennedy}, {Kreykenbohm}, {Kuvvetli}, {Labanti}, {Latronico}, {Li}, {Li}, {Li}, {Li}, {Li}, {Limousin}, {Liu}, {Liu}, {Lu}, {Luo}, {Macera}, {Malcovati}, {Martindale},
  {Michalska}, {Meng}, {Minuti}, {Morbidini}, {Muleri}, {Paltani}, {Perinati}, {Picciotto}, {Piemonte}, {Qu}, {Rachevski}, {Rashevskaya}, {Rodriguez}, {Schanz}, {Shen}, {Sheng}, {Song}, {Song}, {Sgro}, {Sun}, {Tan}, {Uttley}, {Wang}, {Wang}, {Wang}, {Wang}, {Wang}, {Wang}, {Watts}, {Wen}, {Wilms}, {Xiong}, {Yang}, {Yang}, {Yang}, {Yu}, {Zhang}, {Zampa}, {Zampa}, {Zdziarski}, {Zhang}, {Zhang}, {Zhang}, {Zhang}, {Zhang}, {Zhang}, {Zhang}, {Zhang}, {Zhao}, {Zheng}, {Zhou}, {Zorzi}, \& {Zwart}}]{Zhang2019SCPMA}
{Zhang}, S., {Santangelo}, A., {Feroci}, M., {et~al.} 2019, Science China Physics, Mechanics, and Astronomy, 62, 29502, \dodoi{10.1007/s11433-018-9309-2}

\bibitem[{{Zhang} {et~al.}(2022){Zhang}, {M{\'e}ndez}, {Garc{\'\i}a}, {Zhang}, {Karpouzas}, {Altamirano}, {Belloni}, {Qu}, {Zhang}, {Tao}, {Zhang}, {Huang}, {Kong}, {Ma}, {Yu}, {Rawat}, \& {Bellavita}}]{Zhang2022MNRAS}
{Zhang}, Y., {M{\'e}ndez}, M., {Garc{\'\i}a}, F., {et~al.} 2022, \mnras, 512, 2686, \dodoi{10.1093/mnras/stac690}

\bibitem[{{Zhang} {et~al.}(2023){Zhang}, {M{\'e}ndez}, {Garc{\'\i}a}, {Altamirano}, {Belloni}, {Alabarta}, {Zhang}, {Bellavita}, {Rawat}, \& {Ma}}]{Zhang2023MNRAS}
---. 2023, \mnras, 520, 5144, \dodoi{10.1093/mnras/stad460}

\bibitem[{{{\.Z}ycki} {et~al.}(1999){{\.Z}ycki}, {Done}, \& {Smith}}]{zycki1999MNRAS}
{{\.Z}ycki}, P.~T., {Done}, C., \& {Smith}, D.~A. 1999, \mnras, 309, 561, \dodoi{10.1046/j.1365-8711.1999.02885.x}

\end{thebibliography}
\bibliographystyle{aasjournal}

%% This command is needed to show the entire author+affiliation list when
%% the collaboration and author truncation commands are used.  It has to
%% go at the end of the manuscript.
%\allauthors

%% Include this line if you are using the \added, \replaced, \deleted
%% commands to see a summary list of all changes at the end of the article.
%\listofchanges

\end{document}